\documentclass[aps,pra,amsmath,amssymb,amsfonts,%
twocolumn,showpacs,longbibliography]{revtex4-2}

\usepackage{orcidlink}
\usepackage[T1]{fontenc}
\usepackage{graphicx}
\usepackage{dcolumn}
\usepackage{bm}
\usepackage{braket}
\usepackage{amsmath}
\usepackage{mathtools}
\usepackage{graphicx}
\usepackage{xcolor}
\usepackage{multirow}

\usepackage{hyperref}
\hypersetup{colorlinks=true,%
linkcolor=blue!45!black,%
citecolor=blue!45!black,%
urlcolor=blue!45!black, pdfborder={0 0 0}, pdfhighlight=/N}

\usepackage[capitalize]{cleveref}

\renewcommand{\Re}{{\mathrm{Re}}}
\renewcommand{\Im}{{\mathrm{Im}}}

\newcommand{\calN}{{\mathcal{N}}}
\newcommand{\CM}{{\mathrm{CM}}}
\newcommand{\rel}{{\mathrm{rel}}}

\newcommand{\calM}{{\mathcal{M}}}

\newcommand{\dd}{{\rm d}}
\newcommand{\ii}{{\rm i}}
\newcommand{\ee}{{\rm e}}

\newcommand{\LL}{{\mathrm{L}}}

\newcommand{\abs}[1]{\left| #1 \right|} 
\newcommand{\vvv}[1]{{\bm  #1}} 

\newcommand{\rdme}[2]{\left|\langle #1\Vert e\mathbf r\Vert #2\rangle\right|}

\newcommand{\Sr}{^{88}\mathrm{Sr}}

\newcommand{\Rhat}{\widehat{\bm R}}
\newcommand{\Sigmas}{\Sigma_{\!s}}
\newcommand{\Iop}{\widehat{\mathcal I}}

\definecolor{garrosgreen}{rgb}{0.1, 0.4, 0.1}
\definecolor{dartmouthgreen}{rgb}{0.05, 0.5, 0.06}
\definecolor{duelferred}{rgb}{0.7, 0.2, 0.1}
\definecolor{cambridgeblue}{rgb}{0.1, 0.3, 1.0}
\definecolor{oxfordblue}{rgb}{0.05, 0.2, 0.7}

\begin{document}

\title{Quantum control of chemical bonding and antibonding: Entanglement\\
of a single alkaline-earth-metal long-range dimer}

\author{G. A. Bougas \orcidlink{0009-0002-5643-9293}}
%\email{gbougas@mst.edu}
\affiliation{Department of Physics and LAMOR, 
Missouri University of Science and Technology, Rolla, Missouri 65409, USA}

\author{U. D. Jentschura \orcidlink{0000-0003-4967-5458}}
%\email{ulj@mst.edu}
\affiliation{Department of Physics and LAMOR, 
Missouri University of Science and Technology, Rolla, Missouri 65409, USA}

\author{S. I. Mistakidis \orcidlink{0000-0002-5118-5792}}
%\email{smystakidis@mst.edu}
\affiliation{Department of Physics and LAMOR, 
Missouri University of Science and Technology, Rolla, Missouri 65409, USA}

\begin{abstract}
Two identical neutral atoms, one of which is in an excited internal state,
exhibit suppressed and enhanced collective spontaneous emission when prepared
in subradiant and superradiant Dicke states, respectively. A single atom pair
initialized in such a highly entangled state interacts via long-range dipolar
forces, arising from resonant virtual-photon exchange processes. These states
have been studied in long-range alkali-metal and long-range alkaline-earth-metal
dimers in both bonding as well as antibonding orbitals. We here show that a
study of these dimers at the level of a single atom pair (a single molecule)
offers possibilities to gain insights into the dynamics of subradiant (dark)
Dicke states, including the antibonding subradiant longitudinally polarized
Dicke state. We also propose to measure the repulsive forces at the quantum
level directly, by monitoring the resultant atomic trajectories, whose
expectation values follow the Ehrenfest theorem. For the attractive
polarization, we calculate the quantum dynamics of the entangled long-range
dimer prepared in a coherent superposition of long-range bound states. The
detection of the polarization-dependent forces via different cycling
transitions, before and after decay to the ground state, is discussed.
\end{abstract}

\maketitle

\section{Introduction}
\label{Sec:Intro}

The virtual-photon exchange between identical atoms,
one of which is in an excited state,
constitutes a rather peculiar case example of 
interactions between neutral atoms~\cite{Haroche_Nobel,haroche1989cavity}. 
When the atoms exchange virtual photons,
the quantum transition amplitude is enhanced  
at the point of resonance.
Because of the identical character of the two
atoms involved, the quantum state with atom $A$
in the ground state and atom $B$ in the excited
state, is energetically degenerate with the 
quantum state in which 
atoms $A$ and $B$ are interchanged.
Even more interestingly, 
if one entangles the two atoms in a coherent 
superposition of ground-state $A$ and excited-state $B$ and of
ground-state $B$ and excited-state $A$,
then it is possible to either enhance (superradiance)
or suppress (subradiance) spontaneous emission
(with the subradiant state having a minus sign in the superposition;
see Ref.~\cite{craig_molecular_1998}).
In the limit
of close approach of the two atoms, the dipole transition matrix 
element between the subradiant state and the 
ground($A$)-ground($B$) state is vanishing
\cite{footnote}. The subradiant and superradiant states,
whose analytic form is given in Eq.~(\ref{Eq:Dicke}) below,
are known as Dicke~\cite{Dicke_coherence_1954}
or Bell states~\cite{Horodecki_quantum_2009,Nielsen_quantum_2010}.

In general, Dicke states are many-body entangled symmetric or antisymmetric states featuring a fixed number of atomic
excitations~\cite{Prevedel_experimental_2009,%
Stockton_characterizing_2003}.
Due to their properties, Dicke states are paramount in quantum interferometry~\cite{Holland_interferometric_1993,Lucke_twin_2011}.
They also find
applications in quantum information~\cite{Pezze_metrology_2018} and quantum
computing~\cite{Mukherjee_preparing_2020,Kerzner_simple_2025,Shintaro_concrete_2023},
and have been extensively employed in ultracold
gases~\cite{luo2017deterministic,zou2018beating,Pezze_metrology_2018,%
Rui_subradiant_2020,Glicenstein_probing_2025,Takasu_controlled_2012,Guerin_subradiance_2016}. 
The latter
offer the ability to harness quantum matter with unprecedented
applications in quantum science and
technology~\cite{Kaufman_quantum_2021,Altman,cornish2024quantum}.  
Further applications range from quantum metrology~\cite{zhao2021creation}, information
processing~\cite{henriet2020quantum}, and precision
measurements~\cite{Marti,ludlow2015optical} to scalable
architectures~\cite{Altman}. 
A broad set of tools is currently available
for implementing these scalable
architectures, ranging from optical
tweezers~\cite{Kaufman_quantum_2021} and
lattices~\cite{Gross_quantum_2017,Lewenstein_ultracold_2007} in the
subwavelength
regime~\cite{Wang_dark_2018,Yi_state_2008,%
Ritt_Fourier_2006,Tudela_subwavelength_2015} to digital micromirror devices~\cite{Gauthier_direct_2016,Mazurenko_cold_2017}
and high-resolution microscopy
techniques~\cite{Santra_scanning_2015,McDonald_superresolution_2019,%
Subhankar_nanoscale_2019,Bakr_quantum_2009}.
These technological advances share the premise of controllably addressing
large atom arrays as well as individual atoms.

Collective spontaneous emission and subradiance in Dicke states
have been explored across a wide range of platforms, including structured
reservoirs~\cite{Tudela}, band-edge physics~\cite{Stewart}, and
retardation-induced non-Markovian waveguide dynamics~\cite{Sinha}.
Accordingly, subradiance is an established tool for engineering long-lived excitations and controlling light–matter coupling, while recent experiments demonstrated that collective effects may lead to vacuum-induced coherences and directional emission, even in dilute atomic ensembles~\cite{Solano,Han}. 
In the aforementioned studies, the focus has been 
on the modification of radiative decay properties
due to collective entanglement rather than the
entanglement-dependent interaction potentials 
and corresponding forces, which is the focus of our work.

The resonantly enhanced 
interatomic forces studied here are parametrically enhanced 
in comparison to ``usual'' van der Waals forces 
($1/R^3$ for close approach as opposed to $1/R^6$) 
due to the exchange of virtual,
resonant photons between the identical
entangled atoms~\cite{Jentschura_atoms_2023,Salam_Molecular_2009}.
The frequency of the resonant virtual photons equals
the atomic transition frequency of a dipole-allowed
transition. We investigate the singlet $S$ to singlet $P$ 
transition ($5 {}^1 S_0 \to 5 {}^1 P_1$)
in a spinless strontium isotope ($^{88}$Sr atom). 
Thus, we explore possibilities to
study the quantum dynamics of a single alkaline-earth-metal entangled atom pair encoded in a
subradiant Dicke state.
Normally, long-range dimers are created, in large numbers, by
photoassociation inside traps (for an overview,
see Ref.~\cite{Jones2006}). Here we focus
on a single atom pair (a single molecule; see Appendix~\ref{appa}), 
inspired by recent progress in this direction~\cite{Weyland_correlations_2021}.

Our proposal focuses on the actual repulsive and 
attractive forces between the interacting atoms, 
rather than the study of the vibrational 
sublevels of the molecular bound states. In order 
to achieve polarization-selective excitation, 
we assume that the preparation of the 
entangled states proceeds in subwavelength optical lattices, 
with the atoms being subsequently released from their traps;
in a first step, one might also prepare the 
entangled states based on a single atom pair 
in a steep elongated dipole trap (see also Appendix~\ref{appb}).
A detection of the entanglement- and polarization-dependent forces
would deepen our understanding on resonant photon exchange.
In a wider context, such forces 
could be important for engineering 
entanglement-controlled interactions that 
lead to advanced quantum
technologies with atomic arrays and programmable 
quantum simulators~\cite{carr2009cold,softley}. 

This work is structured as follows. In Sec.~\ref{Sec:Setup} details are
provided regarding the resonant dipole-dipole interatomic potentials, the
entangled Dicke states, and the entanglement procedure. In
Sec.~\ref{Sec:Hamiltonians} the nuclear Hamiltonian is derived, and a
center-of-mass transformation is applied to study the relative atomic motion.
In Sec.~\ref{Sec:Quantum_dynamics} we consider the wave packet
dynamics of atoms prepared in either attractive or repulsive Dicke states. In
Sec.~\ref{Sec:Detection} we propose two distinct detection schemes based on
fluorescence imaging. Section~\ref{Sec:Conclusions} summarizes our results and
elaborates on future extensions of our work. In Appendix~\ref{appa} the
connection between Dicke states and molecular spectroscopic notation is
highlighted. Furthermore, Appendix~\ref{appb} discusses different trapping
potentials that can be used for the Dicke state preparation.

\begin{figure}[t!]
\begin{center}
\includegraphics[width=0.99\linewidth]{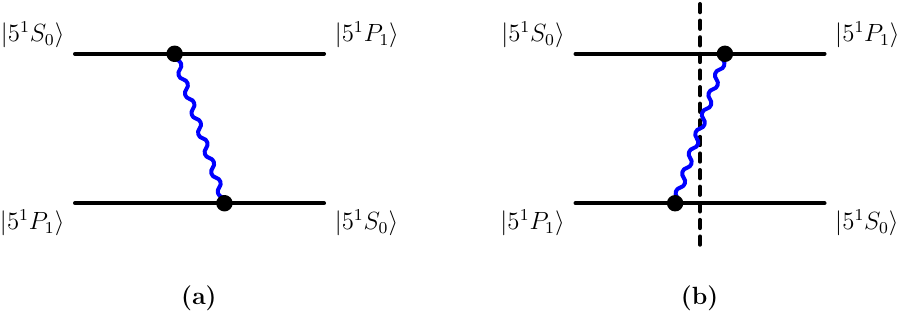}
\end{center}
\caption{\label{figS1}
Two time-ordered (Feynman) diagrams
contribute to the resonant virtual-photon exchange between the ground state
$| g \rangle = | 5 {}^1 S_0 \rangle$
and the excited state $| e \rangle = | 5 {}^1 P_1 \rangle$.
Solid lines denote active electrons, while  
the wavy lines represent  the virtual photon.
In diagram (a), the virtual-photon 
emission proceeds from the ground state. 
In the intermediate atomic state between photon emission
and absorption, both atoms are in the 
$| 5 {}^1 P_1 \rangle$ state. This leads to a purely 
real interaction potential.
In the intermediate atomic state of diagram (b), 
both atoms are in the 
$| 5 {}^1 S_0 \rangle$ state. In 
diagram (b), using ``particle-physics jargon'',
the virtual photon can become an on-shell particle,
which leads to an imaginary part of the energy
shift, by the Cutkosky rules (see Refs.~\cite{Berestetskii_quantum_1982,Cu1960}). 
The intermediate (virtual)
state in diagram (a) has both atoms in the excited state,
while they are both in the ground state in diagram (b).
This explains why the generation of the imaginary 
part is tied to diagram (b).}
\end{figure}

\section{Resonant Dipole--Dipole Interactions}
\label{Sec:Setup}

\subsection{Perturbation theory and matching}
\label{explanation}

Here, we attempt to provide a nonstandard perspective 
on the derivation of the retarded, resonant virtual-photon 
exchange between two identical atoms, one of which is in 
an excited state. We use the formalism of quantum 
electrodynamics and
consider two identical atoms, $A$ and $B$, of atomic mass $M$, in their ground 
($g$) and excited ($e$) states, $\psi_{k,g} \equiv \psi_{k,g}(\vvv{r}_i)$, 
and $\psi_{k,e} \equiv \psi_{k,e}(\vvv{r}_i)$, where $k = A, B$.
The relative coordinate between the nucleus and
the active electron of the $k$th atom is 
denoted by $\vvv{r}_k = x^i_k \, \hat{\vvv{e}}_i$,
where the Cartesian unit vectors are denoted by $\hat{\vvv{e}}_i$,
with $i=1,2,3$, and the Einstein summation convention is used.
The excited state is characterized by a natural decay width 
$\Gamma_0$, and an associated lifetime, $\tau = \hbar / \Gamma_0$. 
Moreover, the resonance (angular)
frequency is $\omega_0 = \abs{E_g-E_e}/\hbar$, where $E_g$ ($E_e$) denotes the 
ground-state (excited-state) energy.

The resonant exchange of virtual photons of frequency $\omega_0$
(see Refs.~\cite{Jentschura_atoms_2023,Salam_Molecular_2009})
couples the states $\psi_{A,g} \, \psi_{B,e} \leftrightarrow \psi_{A,e} \, \psi_{B,g}$.
Matching the $S$-matrix up to second order with an effective
Hamiltonian in a fully quantum-electrodynamic
formalism~\cite{Jentschura_atoms_2023}, 
one obtains an effective Hamiltonian. The general idea behind the matching
of a scattering amplitude ($S$-matrix element) and an effective Hamiltonian
is described in Ref.~\cite{Jentschura_quantum_2022}. The procedure is as follows: $S$-matrix elements are calculated
between in and out states of a physical system,
{\em i.e.}, they describe the 
time evolution of a quantum dynamical system.
Conversely, the time-ordered exponential 
of the interaction Hamiltonian does the same thing,
{\em i.e.}, it describes the evolution of a 
quantum dynamical system from the distant past to the 
distant future.
Hence, an $S$-matrix element and an ``effective''
interaction Hamiltonian, in a given order 
of perturbation theory, can be matched.
Quite famously, for the plain electromagnetic interaction between
two charged spin-$1/2$ particles, this matching procedure has been outlined in Refs.~\cite{Jentschura_quantum_2022,Berestetskii_quantum_1982}.
The leading term (the Coulomb interaction)
and the relativistic corrections (Breit corrections)
are obtained via the matching procedure.
A related question is whether ``effective'' Hamiltonians
could be considered ``fundamental''.
According to the above discussion, a scattering amplitude
and the underlying interaction, physically,
are ``not more and not less fundamental''
than the effective Hamiltonian to which it corresponds.

For illustration, we recall a few details of the
derivation presented in Ref.~\cite{Jentschura_atoms_2023} while temporarily switching to natural units
($\hbar = c = \epsilon_0 = 1$).
For the matching, we consider the action of the potential
$U(\vvv r_A, \vvv r_B, \vvv R)$,
where $\vvv R = \vvv R_A - \vvv R_B$ is the
interatomic distance. The corresponding $S$-matrix element
reads [see also Eq.~(10) of Ref.~\cite{Jentschura_atoms_2023}],
\begin{align}
S_{fi} =& \; - \ii \,
\int \dd^3 r_A \int \dd^3 r_B \;
\psi_A'^*(\vvv r_A) \, \psi_B'^*(\vvv r_B) 
U(\vvv r_A, \vvv r_B, \vvv R) \,
\nonumber\\
& \; \times 
\psi_A(\vvv r_A) \, \psi_B(\vvv r_B) \,
\int \dd t \, \ee^{ -\ii \, (E_A + E_B - E'_A - E'_B) \, t }
\nonumber\\
=& \; -\ii \, T \,
\int \dd^3 r_A \int \dd^3 r_B \;
\psi_A'^*(\vvv r_A) \, \psi_B'^*(\vvv r_B) \,
\nonumber\\
& \; \times U(\vvv r_A, \vvv r_B, \vvv R) \,
\psi_A(\vvv r_A) \, \psi_B(\vvv r_B) \,.
\label{Eq:S_matrix_effective_pot}
\end{align}
Here, we have used energy conservation which translates to the condition 
$E_A + E_B = E'_A + E'_B$.
The (long) time interval for the transition from initial to final state
is defined as $\int \dd t = T$.  For the resonant virtual exchange of photons
between ground-state and excited-state ${}^{88}$Sr
atoms, we have 
$E_A + E_B = E'_A + E'_B = E(5 {}^1S_0) + E(5 {}^1P_1)$.

The first nonvanishing contribution
to the $S$-matrix comes from second-order perturbation
theory, and it reads
[see also Eq.~(16) of Ref.~\cite{Jentschura_atoms_2023}],
\begin{align}
\label{S2ampl}
S^{(2)} =& \; -\frac{1}{2} \int \dd t_1 \int \dd t_2 
\nonumber\\
& \; \times \biggl[ \left< 0 \left|
\mathcal{T}\left[ E^i(\vvv R_A, t_1) \, E^j(\vvv R_B, t_2) \right]
\right| 0 \right> 
\nonumber\\
& \; \quad \times \left< \psi_e \, \psi_g \left|
\mathcal{T} \; d_A^i( t_1) \, d_B^j( t_2)
\right| \psi_g \, \psi_e \right>
\nonumber\\[0.1133ex]
& \; 
+ \left< 0 \left|
\mathcal{T}\left[ E^i(\vvv R_B, t_1) \, E^j(\vvv R_A, t_2) \right]
\right| 0 \right> 
\nonumber\\
& \; \quad \times \left< \psi_e \, \psi_g \left|
\mathcal{T} \; d_B^i( t_1) \, d_A^j( t_2)
\right| \psi_g \, \psi_e \right>
\biggr] \,.
\end{align}
Here, the dipole operators are
$\vvv d_A = e \, \vvv r_A$ and $\vvv d_B = e \, \vvv r_B$,
where the relative active-electron-to-nucleus coordinates 
are denoted as $\vvv r_A$ and $\vvv r_B$ 
for the two identical atoms.
Furthermore, $\mathcal{T}$ refers to  the time ordering operator
for the atomic dipole moments and the 
electric-field operators.
The two terms emerging from the 
time ordering are diagrammatically 
shown in Fig.~\ref{figS1}. For our setup we have
the ground state
$| g \rangle = | 5 {}^1 S_0 \rangle$
and the excited state $| e \rangle = | 5 {}^1 P_1 \rangle$.
During the resonant interaction, 
atom $B$ undergoes a transition from $| e \rangle$ to $| g \rangle$,
while $A$ undergoes a transition from $| g \rangle$ to $| e \rangle$.
In the first expression on the right-hand side of Eq.~\eqref{S2ampl}, the
contribution to the diagram of Fig.~\ref{figS1}(a) corresponds to the term
with $t_1 > t_2$ from the time ordering. In this case, we have emission from
the ground state before absorption from the excited state. In the second
expression on the right-hand side of Eq.~\eqref{S2ampl}, the contribution to
the diagram in Fig.~\ref{figS1}(b) is related to the term with $t_2 > t_1$.
In this case, we have emission from the excited state before absorption from
the ground state.
This sequence is exactly what leads to the 
modification of the imaginary part,
{\em i.e.}, the time-ordered diagram in Fig.~\ref{figS1}(b)
leads to the imaginary part of the interaction Hamiltonian.

After carrying out the integrations in Eq.~\eqref{S2ampl},
one arrives at a compact expression
[see also Eq.~(23) of Ref.~\cite{Jentschura_atoms_2023}],
\begin{multline}
S^{(2)} = -\ii \, T \,
(\omega_0)^2 \, D_F^{ij}( \omega_0, \vvv R) \,
\left< \psi_e \, \psi_g \left| d_A^i \, d_B^j \right|
\psi_g \, \psi_e \right> \,,
\\
\omega_0 = E_e - E_g \,,
\label{Eq:S_matrix_integration}
\end{multline}
with the photon propagator, in the temporal gauge,
given by [see also  Eq.~(85.15) of Ref.~\cite{Berestetskii_quantum_1982} and 
Eq.~(18) of Ref.~\cite{Jentschura_atoms_2023}]
\begin{multline}
\label{Dij}
D_F^{ij}(\omega, \vvv R) =
- \left[ \alpha^{ij} + \beta^{ij} \,
\left( \frac{\ii}{|\omega| R} -
\frac{1}{\omega^2 \, R^2} \right) \right] \,
\frac{\ee^{\ii |\omega| R} }{4 \pi R} \,,
\\
\alpha^{ij} = \delta^{ij} - {\hat R}^i \, {\hat R}^j \,,
\quad
\beta^{ij} = \delta^{ij} - 3 {\hat R}^i \, {\hat R}^j \,.
\end{multline}
where $\hat{\vvv R} = \vvv R / R$.
Comparing Eq.~\eqref{Eq:S_matrix_integration} 
with Eq.~\eqref{Eq:S_matrix_effective_pot}, one 
reads off the interaction potential
\begin{multline}
\label{U}
U(\vvv r_A, \vvv r_B, \vvv R) = 
-\omega_0^2 \,
\left[ \alpha^{ij} + \beta^{ij} \,
\left( \frac{\ii}{\omega_0 \, R} -
\frac{1}{\omega_0^2 \, R^2} \right) \right] 
\\
\times \frac{\ee^{\ii \omega_0 \, R} }{4 \pi R} \,
d_A^i \, d_B^j \,.
\end{multline}
The modulus of the (in general complex
rather than real) virtual photon frequency in Eq.~\eqref{Dij}
is $| \omega | = \sqrt{ \omega^2 + \ii \epsilon}$,
where the branch cut of the complex
square root is taken along the positive real axis,
so that ${\rm Im} \, \sqrt{ \omega^2 + \ii \epsilon} \geq 0$.
Here, $\epsilon \to 0^+$ denotes an infinitesimal
positive quantity. This implies that, for real rather than complex
$\omega$, the definition of $|\omega|$ agrees with the
usual one for the modulus of a real number.
For quantum electrodynamic calculations, it is
manifestly necessary to also consider complex values
of $\omega$ (see Ref.~\cite{Jentschura_atoms_2023}).

\color{black}

\subsection{Real and Imaginary Parts}

We now restore physical units of the Système International (SI) and use the projection operators
\begin{equation}
P^{ij}_{\perp} = \delta^{ij} - \frac{R^iR^j}{R^2}  \,,
\qquad
P^{ij}_{\parallel} = \frac{R^i \, R^j}{R^2} \,,
\end{equation}
which project onto the axis transverse
($\perp$) and parallel ($\parallel$) to the internuclear separation
(the $R^i$ with $i=1,2,3$ denote the Cartesian components).
Restoring SI units, 
we can rewrite the interaction potential $U$,
given in Eq.~\eqref{U}, as follows,
\begin{multline}
\label{Uexpressed}
U(\vvv{r}_A, \vvv{r}_B, \vvv{R}) = 
\biggl[ \left( P^{ij}_{\perp} - 2 P^{ij}_{\parallel} \right) \,
\left( 1 - \ii \frac{\omega_0 R}{c}  \right) \\
- P^{ij}_{\perp} \frac{\omega_0^2 \, R^2}{c^2} \biggr] \,
\frac{ \ee^{\ii \omega_0  R / c} e^2  }{4 \pi \epsilon_0 R^3} \,
r_A^i \, r_B^j \,.
\end{multline}
This potential can be divided into a real and an imaginary 
part, according to Eqs.~(25a) and~(25b) of Ref.~\cite{Jentschura_atoms_2023},
\begin{subequations}
\begin{align}
\label{ReU}
& \Re [U(\vvv{r}_A,\vvv{r}_B,\vvv{R})] = \Bigg\{
\frac{3}{4 \xi^3} \left[ (1-\xi^2) \cos\xi +\xi \sin\xi  \right] P^{ij}_{\perp}
\nonumber\\
& \quad \; - \frac{3}{2\xi^3} \left[ \cos\xi +\xi \sin\xi  \right] P^{ij}_{\parallel}
\Bigg\} \frac{e^2 \omega_0^3}{3\pi \epsilon_0  c^3} r^i_A r^j_B \,,
\\[0.1133ex]
\label{ImU}
& \Im [U(\vvv{r}_A,\vvv{r}_B,\vvv{R})] = \Bigg\{
\frac{3}{4 \xi^3} \left[ (1-\xi^2) \sin\xi -\xi \cos\xi  \right] P^{ij}_{\perp}
\nonumber\\
& \quad \; - \frac{3}{2\xi^3} \left[ \sin\xi -\xi \cos\xi  \right] P^{ij}_{\parallel}
\Bigg\} \frac{e^2 \omega_0^3}{3\pi \epsilon_0 c^3} r^i_A r^j_B \,,
\end{align}
\end{subequations}
where $\xi = \omega_0 R/c$ is the dimensionless 
phase accumulated by the virtual photon on its journey
from atom $A$ to atom $B$. 
We now write down the effective interaction between
these atoms, $U = U(\vvv{r}_A,\vvv{r}_B,\vvv{R})$,
where we assume, 
henceforth, that $\vvv{R} = R \, \hat{\vvv{e}}_x$
is the internuclear axis which we orient along the $x$ axis.
In our treatment, we neglect reduced-mass corrections. Specifically,
\begin{subequations}
\label{Eq:Int_potential}
\begin{align}
U(\vvv{r}_A,\vvv{r}_B,\vvv{R}) = & \;
g^{(\perp)}(\xi) \, \mathcal{V}_{\perp}(\boldsymbol{r}_A,\boldsymbol{r}_B)
\nonumber\\
& \; + g^{(\parallel)}(\xi) \,
\mathcal{V}_{\parallel}(\boldsymbol{r}_A,\boldsymbol{r}_B) \label{Eq:Int_potential_a}  \,,
\\
\mathcal{V}_p(\boldsymbol{r}_A,\boldsymbol{r}_B) = & \;
\; P^{ij}_p \frac{e^2 \omega_0^3}{3\pi \epsilon_0  c^3} x^i_A x^j_B, 
\quad p=\parallel,\perp, \\
g^{(\perp)}(\xi) = & \;  \frac{3}{4 \xi^3}
\left[  (1-\xi^2) \, \ee^{\ii \xi} - \ii \xi \ee^{\ii \xi}  \right]
\nonumber\\
=& \; \frac{3}{4 \xi^3} - \frac{3}{8 \xi} - \frac{\ii}{2} + \frac{9 \xi}{32} + 
{\mathcal O}(\xi^2) \,,
\label{Eq:Perp_pot} \\
g^{(\parallel)}(\xi) =& \;  -\frac{3}{2 \xi^3} \,
\left( \ee^{\ii \xi} - \ii \xi \ee^{\ii \xi}  \right)
\nonumber\\
=& \; -\frac{3}{2 \xi^3} - \frac{3}{4 \xi} - \frac{\ii}{2} + \frac{3 \xi}{16} + 
{\mathcal O}(\xi^2) .
\label{Eq:Parr_pot}
\end{align}
\end{subequations}
Here, the $\mathcal{V}_p$ operators describe the 
electronic, and the $g^{(p)}$ functions the nuclear 
(interatomic) part of the Hamiltonian.
The above considerations lead to a unified derivation of both the imaginary
and real parts of the resonant interaction and their correlation,
{\em i.e.}, the unique association of the
polarization of the entangled states, 
their subradiant and superradiant nature, and
their repulsive and attractive character.
In other approaches~\cite{AnSh1987,An1989,DaJeBrAn2003,JeDaAn2004,%
Sa2015,GrJo2016,Sa2018,JoBr2019},
additional considerations are required in order to 
uniquely fix the sign of the imaginary part in relation
to the real part of the interaction.
The polarization-dependent interatomic potential given in
Eq.~\eqref{Eq:Int_potential} scales approximately as $R^{-3}$ 
for short interatomic distances, similarly to the well-known 
long-range dipolar interactions~\cite{chomaz2022dipolar,cornish2024quantum}.

\subsection{Maximally entangled states}

Indistinguishable atoms in coherent superpositions of electronic states,
$c_1 \, \psi_{A,g} \, \psi_{B,e} + c_2 \, \psi_{A,e} \, \psi_{B,g}$, 
are subject to the interaction potential given by 
Eq.~\eqref{Eq:Int_potential}. The eigenstates of the electron operators $\mathcal{V}_p$ are the 
maximally entangled electronic
states, also known as Dicke states~\cite{Dicke_coherence_1954},
which read
\begin{equation}
\label{Eq:Dicke}
\Psi^{(p)}_{\pm} = \;
\frac{\psi_{A,g}(\boldsymbol{r}_A) \, \psi^{(p)}_{B,e}(\boldsymbol{r}_B)}%
{\sqrt{2}} \pm
\frac{\psi^{(p)}_{A,e}(\boldsymbol{r}_A)
\psi_{B,g}(\boldsymbol{r}_B)} {\sqrt{2}} \,.
\end{equation}
Here, the subscript ($+$)
denotes the ``bright'' superradiant state, 
while the ($-$) subscript designates the ``dark'' subradiant state. Also, the 
superscript $(p)$ stands for the polarization of the excited electronic state
relative to the internuclear axis, {\em i.e.},
transverse ($p = \perp$), or parallel ($p = \, \parallel$).
In the short-range limit ($\xi \to 0$), 
the $\Psi_-^{(p)}$ Dicke states with 
parallel (perpendicular) polarization are subject to repulsive (attractive) interactions. 
The respective superradiant states display the opposite behavior.
In this work, we focus on the following states in ${}^{88}$Sr,
\begin{subequations}
\begin{align}
\label{Eq:eg}
| g \rangle =& \; 
| \; \mbox{[Kr]} \; 5s^2 \; 5{}^1S_0 \rangle \equiv | 5S \rangle \,,
\\
| e \rangle =& \; 
| \; \mbox{[Kr]} \; 5s5p \; 5{}^1P^o_1 \rangle \equiv | 5 P \rangle \,.
\end{align}
\end{subequations}

These singlet states implement the nonrelativistic
angular structure, with magnetic projections
$\mu_z = 0$ for the singlet-$S$ state,
and magnetic projections
$\mu_z = -1,0,1$ for the singlet-$P$ state,
and are connected by the blue magneto-optical trap (MOT) transition
at $460.9\,\rm{nm}$~\cite{PuKrKr88Sr}
(resonant frequency $\omega_0 = 2\pi \times 650.5\,\rm{THz}$, and lifetime
$\tau = 5.253(12)\,\rm{ns}$
\cite{PuEtAl2025,PuKrKr88Sr,NISTASD_2024,Heinz_state_2020,%
Nagel_photoassociative_2005}).
Due to the spinless nucleus of ${}^{88}$Sr,
there is no hyperfine structure.
Such a system avoids the complications of so-called
{\em hyperfine spaghetti} (see Ref.~\cite{Heinzen1994}) curves which lead to considerable
complications for dimers with an involved 
hyperfine structure (see 
Fig.~1 of Ref.~\cite{Williams_1996}
and the remarks on p.~377 of Ref.~\cite{Heinzen1994}).

The decomposition in Eq.~\eqref{Eq:Int_potential_a}
is particularly useful
when evaluating the matrix elements of the polarization-dependent dipole
operators, $\mathcal{V}_p$, contained in the interaction potential $U(\vvv{r}_A,\vvv{r}_B,\vvv{R})$.
The expectation values of the above operators with respect to the electronic
Dicke states $\Psi^{(p=\perp,\parallel)}_{\pm}(\vvv r_A, \vvv r_B)$ reproduce
the natural decay width of the electronically excited state, $\Gamma_0$,
\begin{multline}
\braket{ \Psi^{(p)}_{\pm}(\vvv{r}_A,\vvv{r}_B) | \mathcal{V}_p |
\Psi^{(p)}_{\pm}(\vvv{r}_A,\vvv{r}_B) } 
= \pm \frac{e^2 \omega_0^3}{3\pi \epsilon_0  c^3} P^{ij}_p 
\\
\times \braket{ \psi_{A,g} | r_A^i | \psi^{(p)}_{A,e} } \,
\braket{ \psi^{(p)}_{A,e} | r_A^j | \psi_{A,g} } =
\pm \Gamma_0 \,,
\end{multline}
where the plus (minus) sign pertains to the superradiant (subradiant) state.
In order to derive the above result, we have used the fact that the two atoms
are identical.
Moreover, if we consider an arbitrary superposition of electronic states,
$c_1 \psi_{A,g}\psi^{(p)}_{B,e}+c_2\psi^{(p)}_{A,e}\psi_{B,g}$, with
$\abs{c_1}^2+\abs{c_2}^2=1$, then the expectation value of the pertinent dipole
operator $\mathcal{V}_p$ becomes $2 \,\Re(c^* _1 \, c_2) \, \Gamma_0$.

\subsection{Entanglement procedure}
\label{Sec:Entanglement}

It is important that the size of the ${}^{88}$Sr atoms 
is much smaller than the dipole transition wavelength.
This ensures that there is vanishing overlap between the 
quantum atomic-state eigenfunctions of the 
two identical atoms
over the distance range of the dipole transition wavelength,
$\lambda_0 = 2 \pi c/\omega_0$. As a remark,
one can easily show that this condition is not met by (very large) 
Rydberg atoms with exceptionally high 
principal quantum numbers $n \sim 80$.
The atomic wave functions of ${}^{88}$Sr (in $5S$ and $5P$ states)
extend over a few Bohr radii, while 
the interatomic separation, for the 
schemes proposed here, is in the range of a few hundred angstroms.

\begin{figure}[t!]
\centering
\includegraphics[width=0.99\linewidth]{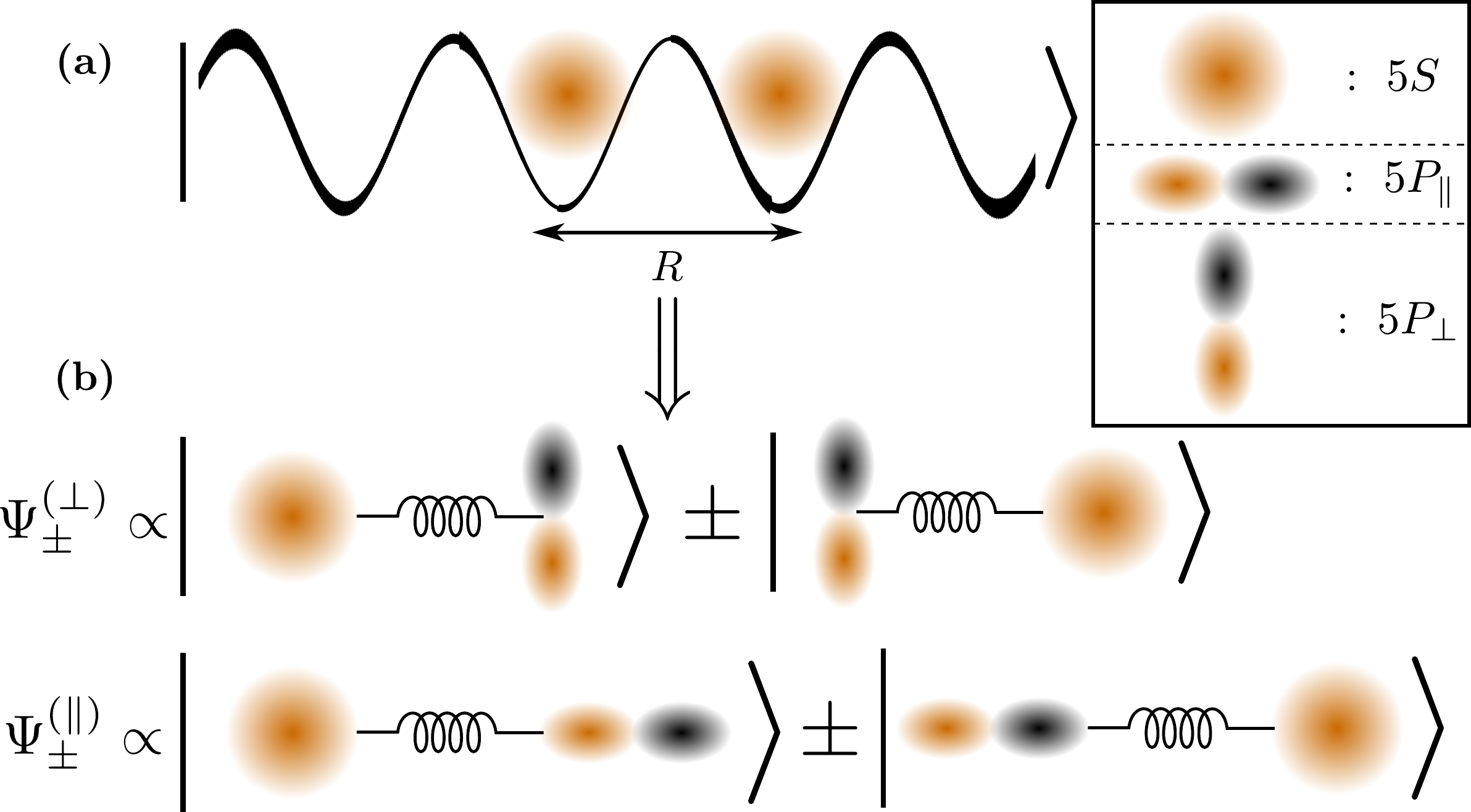}
\caption{Illustration of Dicke--state preparation. (a)
Two identical ${}^{88}$Sr atoms are initially held at a 
distance~$\vvv{R} = R \, \vvv{e}_x$
by means of a subwavelength optical lattice. 
The potential is switched off and the
atoms are prepared in entangled pairs of 
ground and excited states, with the excited state
polarized either along the interatomic 
$x$ axis ($ 5P_\parallel $ state) or across the $y$ or $z$ axis ($5P_\perp $ state).
(b) The Dicke states $\Psi^{(\perp)}_{\pm}$ ($\Psi^{(\parallel)}_{\pm}$),
for transverse (parallel)
polarization are given by Eq.~\eqref{Eq:Dicke}.
The brown (black) color indicates  the positive
(negative) sign of the electronic orbitals, designated in the legend.
The atoms are now interacting with resonant dipole-dipole interactions, 
as schematically denoted by the coil.}
\label{fig1}
\end{figure}

To illustrate the entanglement-dependent, and polarization-dependent,
interactions, we consider two 
initially non-entangled ground-state 
${}^{88}$Sr atoms.
We assume that these are initially trapped by means of subwavelength optical
lattices~\cite{Wang_dark_2018,Tudela_subwavelength_2015} 
(see also Appendix~\ref{appb}) and prepared in their
motional ground states [see also Fig.~\ref{fig1}(a)]. 
Based on the considerations reported in Appendix~\ref{appb},
this represents an experimentally challenging, but
certainly not impossible task given recent experimental advances.
The interatomic distance can be arbitrarily adjusted via, {\em e.g.}, guided modes of photonic
nanocrystals~\cite{Tudela_subwavelength_2015}. Subsequently, we assume that 
entangled electronic Dicke states are prepared (in close range) 
either by means of a sequence of $X$, Hadamard, controlled-$\mathrm{NOT}$ and/or $Z$
gates~\cite{Nourallah_preparation_2023,Mukherjee_preparing_2020,Shintaro_concrete_2023},
or, by polarization-selective photoassociation. In the latter scheme the appropriate 
Condon radius of 50\,nm enables us to select a suitable wavelength
for the laser that operates near the 
free-field blue MOT transition $5{}^1S_0 \to 5{}^1P_1$. 
The latter is, of course,  detuned with the appropriate binding correction
(for the attractive, subradiant state) or the 
appropriate antibonding energy correction
(for the repulsive, subradiant state).

In a wider context, we mention, for the generation of Dicke states,
various other possibilities, such as Laguerre--Gaussian beams that have been
employed in coupled quantum emitters~\cite{Trebbia_tailoring_2022},
single-photon beam splitters~\cite{Scully_single_2015}, and transverse
illumination of atom arrays in cavities~\cite{Gabor_demonstration_2025}.
Additionally, collective subradiant modes have been observed in ultracold
${}^{87}$Rb atoms arranged in subwavelength optical
lattices~\cite{Rui_subradiant_2020,Douglas_many_2026}, using light resonant
with the $D_2$ transition. Either immediately after or  before the entanglement procedure 
the external potential is
switched off, and the atoms experience the interaction potential described by
Eq.~\eqref{Eq:Int_potential}.  Depending on the polarization, different
detection schemes can be employed, all of them within current experimental
reach, see Sec.~\ref{Sec:Detection} for more details.

Owing to the form of the interaction [see
Eq.~\eqref{Eq:Int_potential}], the 
vibrational atomic motion of the long-range dimer
occurs along one spatial direction, 
which we identify as the $x$ axis. 
The spacing between the rotational energy levels corresponding to
motion about the $y$ and $z$ axes 
is much less pronounced as compared to vibrational motions.
(Of course, the entire long-range dimer will still experience
acceleration due to gravity, which becomes relevant 
once the potential is switched off).
An analogous pattern (confined motion perpendicular
to a defined axis, combined with nearly free quantum motion
along the symmetry axis) is observed for so-called tubular image states 
which are related to the geometry of carbon
nanotubes~\cite{Granger_highly_2002,Segal_tunable_2005,Hofer_time_1997,Zamkov_time_2004}.

\section{Nuclear Hamiltonian}  \label{Sec:Hamiltonians}

\subsection{Born-Oppenheimer approximation} 
\label{Sec:Nuclear_Hamilt}

The Hamiltonian pertaining to both atoms,  after the entanglement procedure, can
be written as
\begin{equation}
\label{Eq:Overall_Hamilt}
H = H_A(\vvv{r}_A,\vvv{X}_A) + 
H_B(\vvv{r}_B,\vvv{X}_B) + 
U(\vvv{r}_A,\vvv{r}_B,\vvv{R}).
\end{equation}
Here, $H_i$, $i=A,B$, is the Hamiltonian of the $i$th atom, while $\vvv X_i$ is
the center-of-mass coordinate of the $i$th nucleus with respect to its active
electron.  Up to corrections of order $m_e/M$, where $m_e$ is the electron mass
and $M$ is the mass of the nucleus, the center-of-mass coordinates $\vvv{X}_i$
correspond to the coordinates of the respective nuclei. We work in the
non-recoil approximation from now on.  Furthermore, we assume that the
Hamiltonians $H_A$ and $H_B$ include a decay-rate operator describing the
dipole decay $| e \rangle \to | g \rangle$, of the form
\begin{equation}
-\ii \, \frac{\Gamma_0}{2} \, | e \rangle \, \langle e |  \,,
\end{equation}
which shifts the energy of the excited state according to 
$E_e \to \Re \, [E_e] - \ii \, \Gamma_0/2$. 
We now employ the Born--Oppenheimer
approximation~\cite{Sutcliffe_quantum_2012}.
After the Dicke-state generation, 
the wave function of the two atoms takes the form
\begin{gather}
\Phi_{\pm}^{(p)}(\vvv{r}_A,\vvv{r}_B,\vvv{X}_A,\vvv{X}_B) 
= \Psi^{(p)}_{\pm}(\vvv{r}_A,\vvv{r}_B) \;
\chi^{(p)}_{\pm}(\vvv{X}_A,\vvv{X}_B) \,,
\label{Eq:Entangled_state}
\end{gather}
where the $\chi$ wave function takes into account the nuclear part, while
$\Psi^{(p)}_{\pm}$ corresponds to the entangled state pertaining to the
electronic degrees-of-freedom [see Eq.~\eqref{Eq:Dicke}].

Our next task is to separate the nuclear from the electronic degrees of freedom, 
which will enable us to analyze the quantum dynamics 
of the center-of-mass coordinates of the two atoms. Specifically, by taking 
the expectation values of the Hamiltonian $H$ in the 
space of entangled Dicke states given by  
Eq.~\eqref{Eq:Entangled_state}, it is possible to show that one can 
replace, in Eq.~\eqref{Eq:Overall_Hamilt},
the potential $U(\vvv{r}_A, \vvv{r}_B, \vvv{R})$
by the $\vvv{r}_k$-independent ($k=A,B$) expressions, i.e.
\begin{subequations}
\begin{align}
U(\vvv{r}_A, \vvv{r}_B, \vvv{R}) \to & \;
V^{(p)}_{\pm}(\vvv{R}) - 
\frac{\ii}{2} \, \Gamma_{\pm}^{(p)}(\vvv{R}) \,,
\\
V^{(p)}_{\pm}(\vvv{R}) =& \; \pm \Gamma_0 \, \Re[g^{(p)}(\vvv{R})] \,,
\\
\Gamma^{(p)}_{\pm}(\vvv{R}) =& \; \mp 2 \Gamma_0 \, \Im[g^{(p)}(\vvv{R})] \,, 
\end{align}
\end{subequations}
see also Eqs.~\eqref{Eq:Perp_pot} and~\eqref{Eq:Parr_pot}.
Therefore, as a function of the electronic-state polarization $p$ and the
superposition sign $\pm$ of the Dicke state, the nuclear motion is dictated
by the effective Hamiltonian
$\mathcal{H}^{(p)}_{\pm}(\vvv{X}_A,\vvv{X}_B,\vvv{R})$, which reads 
\begin{multline}
\mathcal{H}^{(p)}_{\pm}(\vvv{X}_A,\vvv{X}_B,\vvv{R}) = 
-\frac{\hbar^2}{2 M} \vvv{\nabla}_{\vvv{X}_A}^2 - 
\frac{\hbar^2}{2 M} \vvv{\nabla}_{\vvv{X}_B}^2 + E_g 
\\
+ \Re \, [E_e] 
+ V^{(p)}_{\pm}(\vvv{R}) -
 \frac{\ii}{2} \,
\left[ \Gamma_0  + \Gamma^{(p)}_{\pm}(\vvv{R})  \right]. 
\label{Eq:Nuclear_Hamilt}
\end{multline}
Here, we have effectively integrated out the electron motion from the Schr\"odinger
equation pertaining to the Hamiltonian of Eq.~\eqref{Eq:Overall_Hamilt}.
In our treatment, we have assumed that the electron-nucleus 
distances, $| \vvv r_A |$ and $| \vvv r_B | $,
are much smaller than the internuclear distance $| \vvv R|$.
This assumption is justified if
the internuclear separation $| \vvv{R} |$ is initially made much
larger than the average size of the atoms by means of subwavelength optical lattices.

\subsection{Center-of-mass transformation} 
\label{Sec:CM_transformation}

To probe the dynamical behavior of the two atoms,
we perform a transformation to
their center-of-mass frame, 
\begin{equation}
\vvv{X}_{\rm CM} = \frac{\vvv{X}_A + \vvv{X}_B}{2}, 
\qquad 
\vvv{R} = \vvv{X}_A - \vvv{X}_B \, .
\label{Eq:COM_transformation}
\end{equation}
Due to the form of the interaction potential [see also
Eq.~\eqref{Eq:Nuclear_Hamilt}] the center-of-mass motion decouples from the
relative motion. Therefore, $\mathcal{H}^{(p)}_{\pm}$ can be written 
as the sum of a center-of-mass Hamiltonian
$\mathcal{H}_{\CM}$ and a relative-frame nuclear Hamiltonian
$\mathcal{H}^{(p)}_{\rel,\pm}$, as follows 
\begin{subequations}
\begin{align}
\label{Eq:COM_nuclear} 
\mathcal{H}_{\CM} = & \; -\frac{\hbar^2}{2 \calM}
{\bm \nabla}^2_{\vvv{X}_{\CM}} + E_g + \Re \, E_e  \,,
\\
\label{Eq:Rel_nuclear}
\mathcal{H}^{(p)}_{\rel,\pm} =& \; 
- \frac{\hbar^2}{2\mu} {\bm \nabla}^2_{\vvv{R}} -\frac{\ii}{2} 
\left[ \Gamma_0 + \Gamma^{(p)}_{\pm}(\vvv{R}) \right] + 
V^{(p)}_{\pm}(\vvv{R}) \,.
\end{align}
\end{subequations}

Here, $\calM=2M$ is the total mass of the two-atom
system, and $\mu = M/2$ is the two-body reduced mass.
For the numerical calculations, we use the 
well-known $^{88}\mathrm{Sr}$
atomic mass $M=87.9056\,\mathrm{u}\simeq
1.4597\times10^{-25}\,\mathrm{kg}$, so that
$\mu=M/2\simeq7.2985\times10^{-26}\,\mathrm{kg}$.
Both functions $g^{(p)}(\vvv{R}) = g^{(p)}(R) = g^{(p)}(\xi)$
with $p = \perp, \parallel$ have the
property $\lim_{R \to 0} \Im[g^{(p)}(R)] = -1/2$. In the case of the
subradiant Dicke state, the effective decay width
$\Gamma_0 +  \Gamma^{(p)}_-(\vvv{R})$
vanishes in the short-range limit $R \to 0$.
The decoupling of the Hamiltonians,
\begin{equation}
\mathcal{H}_{\CM} = \mathcal{H}_{\CM}(\vvv X_{\CM}) \,,
\qquad
\mathcal{H}^{(p)}_{\rel, \pm} = \mathcal{H}^{(p)}_{\rel, \pm}(\vvv R) \,,
\end{equation}
translates also to the nuclear wave functions, {\em i.e.},
\begin{equation}
\chi^{(p)}_{\pm}(\vvv{X}_A,\vvv{X_B}) = 
\chi_{\CM}(\vvv{X}_{\CM}) \, \chi^{(p)}_{\rel,\pm}(\vvv{R}) \,, 
\end{equation}
where $\chi_{\CM}$ ($\chi^{(p)}_{\rel,\pm}$) refers to the center-of-mass (relative) motion.
Within the space of entangled states
considered here [see also Fig.~\ref{fig1}], $\chi_{\CM}$ is a Gaussian wave packet, propagating freely in time.
On the other hand, the evolution of the wave packet in the 
relative coordinate $\boldsymbol{R}$ is governed by the potential
$ V^{(p)}_{\pm}(\vvv{R})$.  

As argued above, the center-of-mass motion is almost free, i.e., it is
governed by the free Schr\"odinger equation and the acceleration due to
gravity, which could otherwise be incorporated by adding a term $\mathcal{M}gz$ to the
Hamiltonian, where $g$ is the acceleration due to gravity and $z$ is the height
coordinate.
Furthermore, under this approximation,
the dynamics of the two atoms along $x$ is  determined by the Schr\"odinger equation associated with
the relative Hamiltonian [Eq.~\eqref{Eq:Rel_nuclear}],
\begin{multline}
\Bigg \{ -\frac{\hbar^2}{2\mu} \frac{\partial^2}{\partial x^2} - 
\frac{\ii}{2} \left[ \Gamma_0  + \Gamma^{(p)}_{\pm}(x)  \right] 
+ V^{(p)}_{\pm}(x) \Bigg\} \, \chi^{(p)}_{\rel,\pm}(x,t) 
\\
= \ii \hbar \, \frac{\partial \chi^{(p)}_{\rel,\pm}(x,t)}{\partial t} \,.
\label{Eq:Rel_nuclear_1D}
\end{multline}
This equation is used here
to investigate the dynamics induced by the 
resonant virtual-photon exchange between the two identical atoms
in the subradiant Dicke states.

\begin{figure}[t!]
\centering
\includegraphics[width=1\columnwidth]{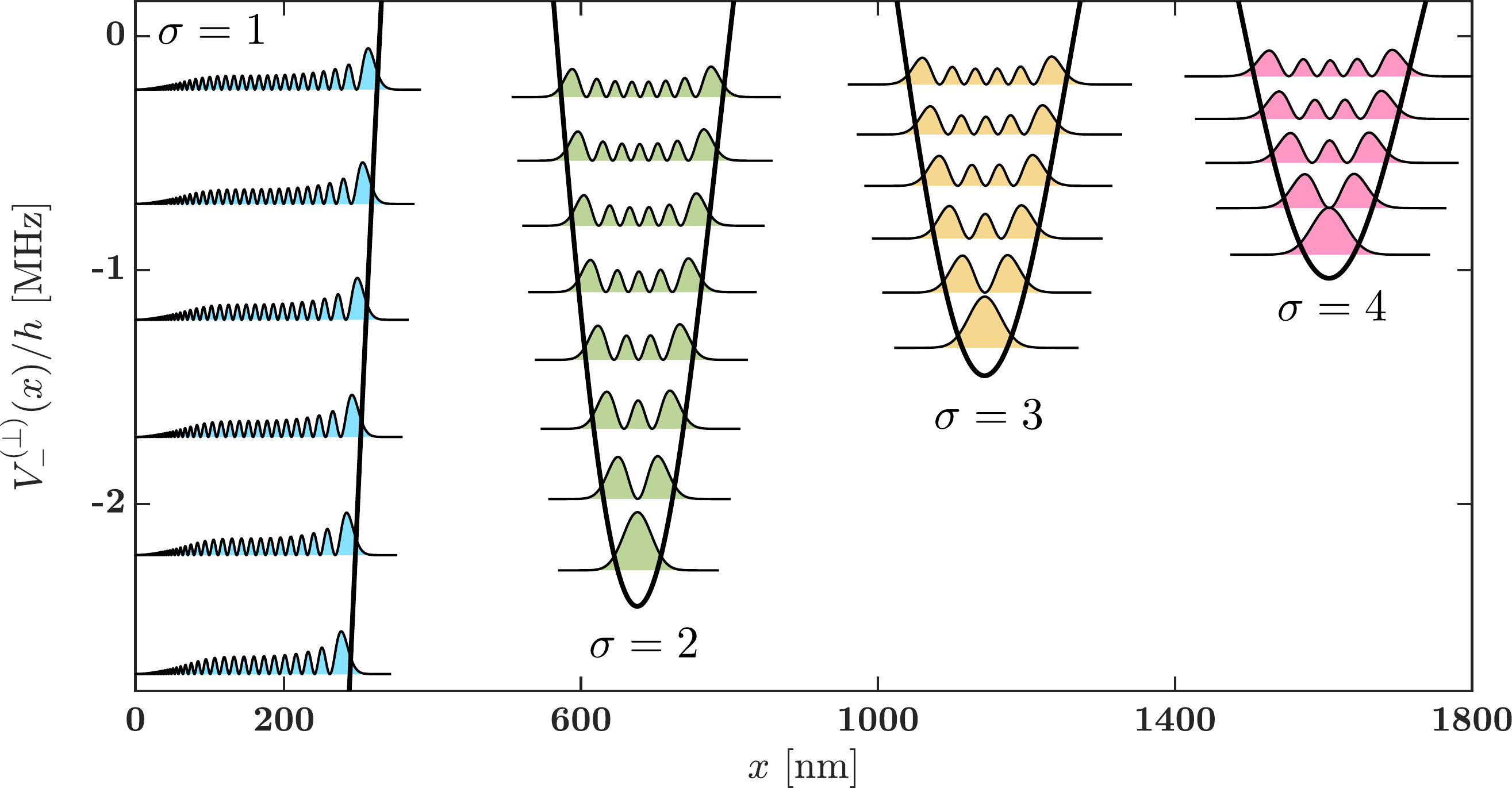}
\caption{
The first few quasimolecular states, $\abs{\varphi_{\bar{n}_{\sigma}}(x)}^2$, 
existing in the first four wells ($\sigma=1,\ldots,4$) of the 
$V^{(\perp)}_-(x)$ potential of ${}^{88}$Sr atoms. The $\bar{n}_{\sigma}$ eigenstates 
are strictly accommodated  within their respective $\sigma$ potential well. 
The majority of the bound states are hosted in the $\sigma=1$ well, 
while the fewer bound states in the outer wells 
are localized at large interatomic distances.}
\label{Fig:Eigenstates}
\end{figure}

\section{Quantum Dynamics of Subradiant States}  \label{Sec:Quantum_dynamics}

\subsection{Subradiant quasimolecular states}
\label{Sec:Eigenstates}

For the perpendicular polarization
[see Eq.~\eqref{Eq:Int_potential}], the subradiant states,
which are the focus of our investigations here,
are the attractive ones (see also Appendix~\ref{appa} for their spectroscopic notation). This follows from the fact that
$- \Re[g^{(\perp)}(\xi)]$
has a leading attractive asymptotics for $\xi \to 0$, 
according to Eq.~\eqref{Eq:Perp_pot}.
This attractive potential leads to the confinement of the two
${}^{88}$Sr atoms within small distances, due to the population of weakly bound,
long-range molecular states.  The potential is strongly attractive, scaling as $R^{-3}$ at
short distances, and becoming highly oscillatory asymptotically
[see also Eq.~\eqref{Eq:Int_potential}].
As a result, many local potential wells form
at large interatomic distances, see Fig.~\ref{Fig:Eigenstates}, supporting a
plethora of bound states with very distinct length scales. 

Despite the
complexity of the bound states, it turns out that they are well localized
within the  individual potential wells labeled by $\sigma=1,2,3,\ldots$.  Therefore, the 
eigenstates, $\varphi_{\bar{n}_{\sigma}}(x)$,  accommodated in the $\sigma$ potential well of $V_-^{(\perp)}(x)$,  are classified
according to the quantum numbers $\bar{n}_{\sigma}$, counting from the least
bound state at the zero-energy threshold.  The first potential well hosts most
of the bound states $(\rm{approximately}~170)$, while the outer $\sigma$ wells support a
small number of bound states [see Fig.~\ref{Fig:Eigenstates}]. 
The bound states in the outer wells are localized at
extremely large interatomic distances, and they are irrelevant for the
wave packet propagation.  Instead, only the weakly bound quasimolecular levels of the
$\sigma=1$ well play the major role.

In our numerical simulations, the eigenstates of the $\sigma=1$ well are
evaluated down to interatomic distances $x_c = 1.19~\rm{nm}$. This cutoff
corresponds to the LeRoy radius of two ${}^{88}$Sr
atoms~\cite{Iritani_accurate_2023,LeRoy_dissociation_1970}, one in the ground
state, the other one in the excited $5{}^1P_1$ state with polarization
perpendicular to the internuclear axis. The LeRoy radius provides an estimate
of the distance where the electronic orbitals overlap significantly, leading to
the formation of short-range molecular states. At such short distances the formalism for
the interaction potentials is not valid anymore, and therefore a hard wall is
placed at $x_c$.  This hard wall, however, does not affect the behavior of the
higher-lying eigenstates at large interatomic distances.  The
$\varphi_{\bar{n}_{\sigma}}$ states are expanded in terms of B-spline basis
sets~\cite{Boor_practical_1978} and the expansion coefficients are determined
by the eigenvectors of the relevant matrix
equation~\cite{Bachau_applications_2001} pertaining to the stationary
Schr\"{o}dinger equation [Eq.~(\ref{Eq:Rel_nuclear_1D})] of the
$V^{(\perp)}_{-}(x)$ potential.  The eigenspectrum of the latter provides an
alternative route for propagating the initial wave packet, in good agreement
with the dynamics presented in Fig.~\ref{Fig:Wave_prop}(b).  Within the
decomposition of the initial state into eigenstates, the time evolution simply
amounts to the multiplication of the individual terms in the eigenstate
decomposition by phase factors $\exp(-{\rm i} \, E_{\bar{n}_\sigma} t /\hbar)$. 

\subsection{Wave-packet propagation}  
\label{Sec:Prop}

We assume that, as a result of the 
preparation scheme outlined in Sec.~\ref{Sec:Entanglement}, the  
initial state is a Gaussian wave packet with controllable mean position and
width. These two parameters serve as tunable probes for the interatomic potential of 
Eq.~\eqref{Eq:Int_potential_a}.  The relative motion of the two
atoms subject to $U(\vvv{r}_A,\vvv{r}_B,\vvv{R})$ is
characterized by the wavefunction $\chi^{(p)}_{\rel,\pm}(x,t)$.
Initially, the Gaussian wave packet $\chi^{(p)}_{\rel,\pm}(x,0)$ is 
\begin{equation}
\chi^{(p)}_{\rel,\pm}(x,0) = \calN^{-1/2} \,
e^{- (x-x_0)^2/(2 w^2) } \,,
\end{equation}
where $\calN = w \frac{\sqrt{\pi}}{2} [1 + \text{erf}(x_0/w)]$, 
$\rm{erf}(\cdot)$ is the error function, $x_0$ is the initial interatomic
distance and $w$ is the width of the wave packet.  
This wave packet is {\em not} an 
eigenstate of the Hamiltonian of  Eq.~\eqref{Eq:Rel_nuclear},
and, hence, its time evolution is nontrivial
and it is described by the time-dependent 
Schr\"{o}dinger equation for the nuclear motion
in ``dark'' subradiant states of polarization $p$ [Eq.~\eqref{Eq:Rel_nuclear_1D}].

\begin{figure*}[t!]
\centering
\includegraphics[width=0.99\linewidth]{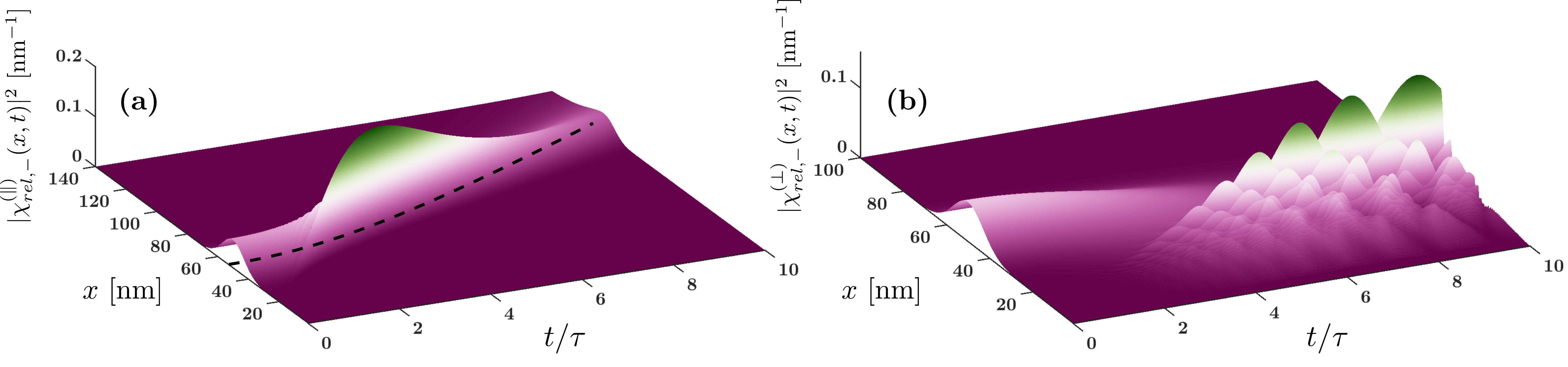}
\caption{
Impact of the polarization-dependent interactions in
the wave-packet propagation of two ${}^{88}$Sr atoms.  The latter are
initialized in the subradiant Dicke state with (a) parallel and (b)
transverse polarization relative to their internuclear axis. The black
dashed line in panel (a) stems from a semiclassical approximation of the mean
position. The Dicke states are prepared by employing the
blue MOT transition $5{}^1S_0 \to 5{}^1P_1$.
The initial state of the two atoms is a Gaussian
wave packet centered at 
$x_0 = 50 \, \mathrm{nm}$ with a width of
$w = 10 \, \mathrm{nm}$.}
\label{Fig:Wave_prop}
\end{figure*}

In the short-range limit $R \to 0$,
the atom-atom interaction potential 
for the parallel-polarized state, $V^{(p=\parallel)}_-$, is strongly repulsive (see also Appendix~\ref{appa} for their spectroscopic notation).
This translates into a repulsive interaction for nuclei in the
$\chi^{(p=\parallel)}_{\rel,-}(x,t)$ state. For
the parameters chosen in Fig.~\ref{Fig:Wave_prop}(a), the repulsion is so
strong that after $t=10~\tau$, the mean position $\braket{x(t)}$ of the
wave packet is increased by a factor of $2.2$.  
The rapid increase in $\braket{x(t)}$ can be
understood by treating the wave packet's center-of-mass as a classical particle.
Its trajectory is determined by the Ehrenfest equation, 
\begin{equation}
\mu \frac{d^2 \braket{x(t)}}{dt^2} = 
- \left< \frac{dV^{(\parallel)}_-(x)}{dx} \right> \,,
\end{equation}
and the solution is shown with the dashed curve in Fig.~\ref{Fig:Wave_prop}(a). 
It becomes evident that it is 
in good agreement with the numerically determined quantum dynamics.

By contrast, the atom-atom interaction potential 
$V^{(p=\perp)}_-$ is strongly attractive
for the transversely polarized
$| \Psi^{(p=\perp)}_- \rangle$ electronic state, which translates 
into the $\chi^{(p=\perp)}_{\rel,-}(x,t)$ state for the motion of the nuclei.
In this case, the quantum dynamical evolution of the 
wave packet, under the attractive potential, implies that the two atoms 
remain trapped within small distances [see Fig.~\ref{Fig:Wave_prop}(b)]. 
For the parameters of Fig.~\ref{Fig:Wave_prop}(b),
the wave packet fragments and develops a multi-peak
structure, which, partially, even recurs in time. 

\begin{figure}[t!]
\centering
\includegraphics[width=0.99\linewidth]{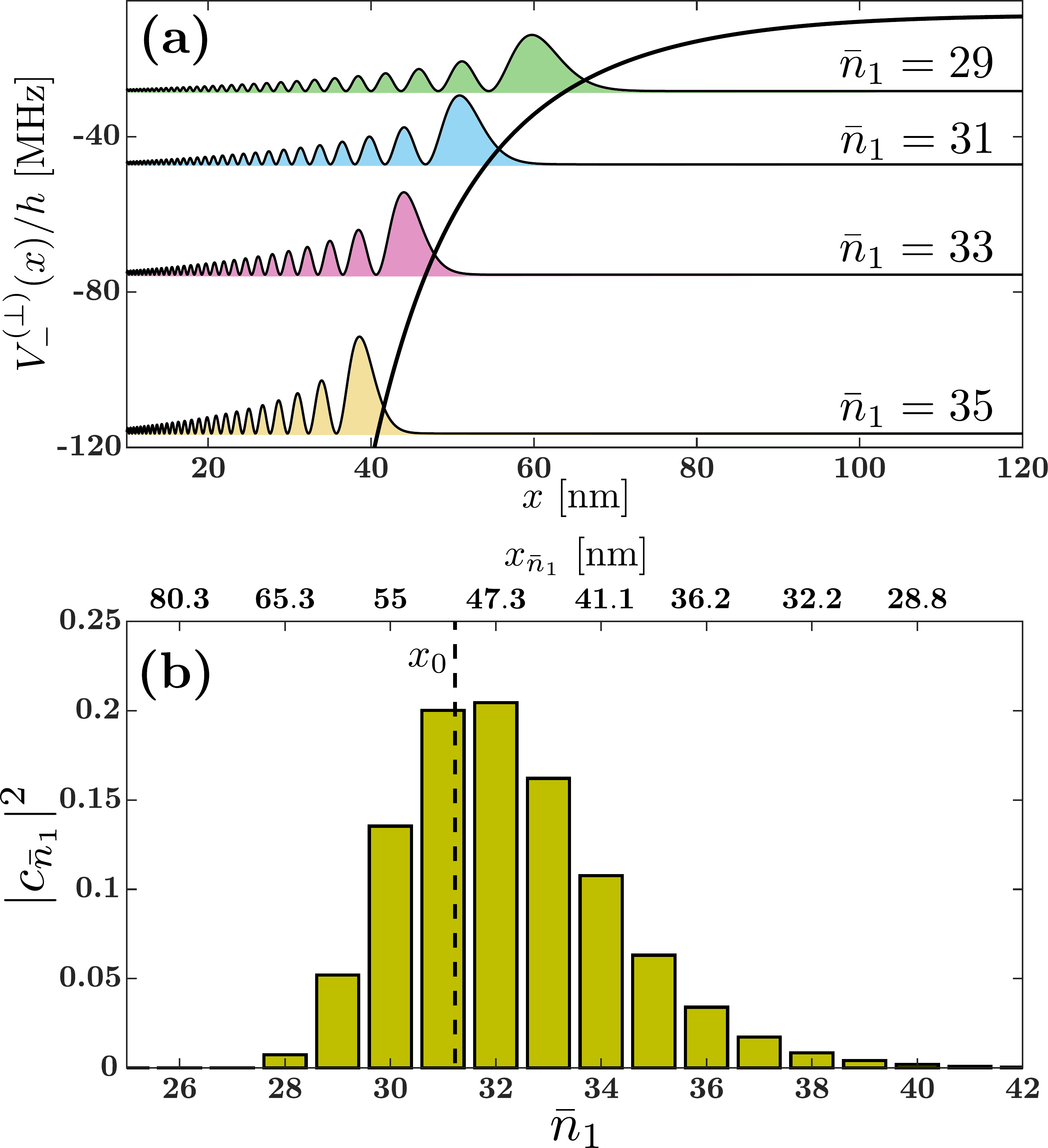}
\caption{\label{Fig:Overlaps}
Selective population of weakly
bound quasimolecular states of the $\sigma=1$ potential well. For the
attractive $V^{(\perp)}_-(x)$ potential, the ${}^{88}$Sr atoms form 
quasimolecular states, a few of them 
depicted in panel (a), along with their quantum numbers, $\bar{n}_1$. 
(b) Weakly bound 
quasimolecular states whose
outer classical turning point, $x_{\bar{n}_1}$, is close to $x_0$ (dashed line) are predominantly contributing, as identified from the eigenstate overlap coefficients 
$\abs{c_{\bar{n}_1}}^2$. The lower and upper abscissas
in panel (b) represent the quantum numbers and classical turning points, respectively.
}
\end{figure}

Such a structure stems from the non-eigenstate 
property of $\chi^{(p=\perp)}_{\rel,-}(x,t)$, which 
has contributions from a large number of bound
quasimolecular eigenstates of $\mathcal{H}^{(\perp)}_{\rel,-}$.
Roughly ten molecular states ($28 \leq \bar{n}_1 \leq 38$) from the
innermost ($\sigma=1$) potential well contribute to the dynamics [see Fig.~\ref{Fig:Overlaps}(a), (b)]. 
The amplitudes with which 
these states contribute to the 
quantum dynamics can be determined through the overlap coefficients 
$c_{\bar{n}_1}$ of the wave packet with
the eigenstates $\varphi_{\bar{n}_1}(x)$ of 
$\mathcal{H}^{(\perp)}_{\rel,-}$, {\em i.e.},
\begin{equation}
c_{\bar{n}_1} = \int \dd x \, 
\varphi^*_{\bar{n}_1}(x) \, \chi^{(\perp) }_{\rel,-}(x,0) \,.
\end{equation}
The absolute square of the expansion coefficients $\abs{c_{\bar{n}_1}}^2$ is
appreciably different from zero only for those eigenstates whose 
outer classical turning point, $x_{\bar{n}_1}$, remains
within the innermost ($\sigma =1$) potential well and is near the
wave packet's mean position,
$x=x_0$ (see Refs.~\cite{Bougas_dynamical_2022,Colussi_dynamics_2018}) as can be seen in 
Fig.~\ref{Fig:Overlaps}(b).

\begin{figure}[t!]
\centering
\includegraphics[width=0.99\linewidth]{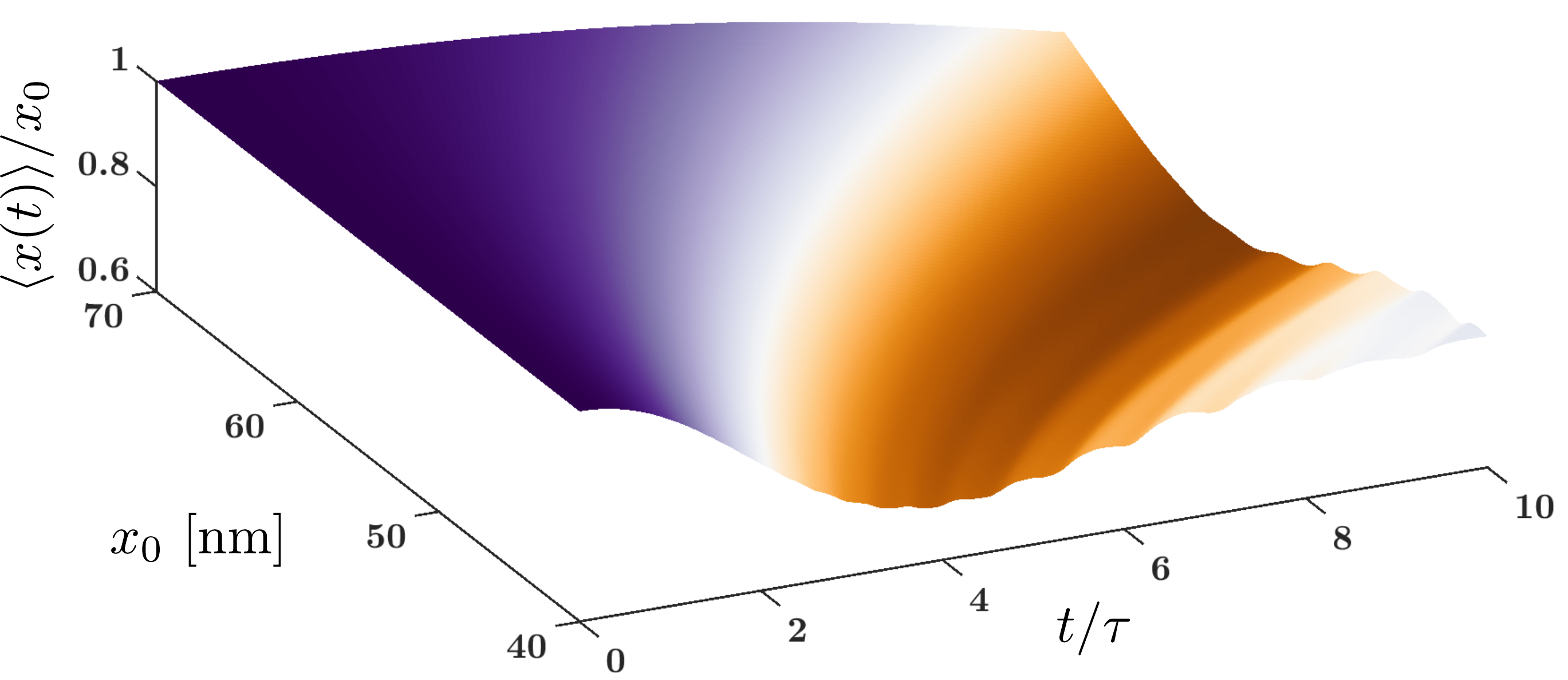}
\caption{ Normalized average distance of two ${}^{88}$Sr atoms,
$\braket{x(t)}/x_0$, when prepared in the 
attractive subradiant $\Psi^{(\perp)}_-$ Dicke state. The
initial {\em width} of the Gaussian 
wave packet is fixed to $w = 10 \, \mathrm{nm}$, while the initial 
{\em mean separation}, $x_0$, is varied. Depending on $x_0$, different 
bound states are populated [see also Fig.~\ref{Fig:Overlaps}(b)],
a behavior manifested by the different frequencies of the 
oscillatory beats of $\braket{x(t)}$.}
\label{Fig:Distance_scan}
\end{figure}

Capitalizing on the tunability of the initial wave packet's position, we can
even control the quantum numbers of the participating eigenstates.  Different
values of $x_0 \equiv  \braket{x(0)}$ lead to the population of distinct
quasimolecular states, whose outer classical turning points are centered
around the mean wave packets position $x_0$ [see also
Fig.~\ref{Fig:Overlaps}(b)]. The involvement of distinct bound states depending
on $x_0$ manifests in the dynamical behavior of the normalized mean position of
the wave packet, $\braket{x(t)}/x_0$, as presented in
Fig.~\ref{Fig:Distance_scan}.  The frequencies of the oscillatory beats of
$\braket{x(t)}/x_0$ (the ``width of the ripples on the surface'') correspond to
the energy gaps of the contributing eigenstates, e.g. between $\bar{n}_1 = 30$
and $\bar{n}_1=34$ for $x_0 = 40 ~ \rm{nm}$.

For relatively small $x_0 < 50 \, \mathrm{nm}$, such states are
characterized by small classical turning points, and thus feature large energy
gaps, since they are deeply bound in the binding potential
$V^{(\perp)}_-(x)$ [see also
Fig.~\ref{Fig:Overlaps}(a)]. Hence, the oscillatory beats
become faster as compared to larger internuclear separations
(the apparent beat frequency is increased,
and the width of the ``ripples on the surface''
during evolution is decreased).

\section{Detection Schemes}  \label{Sec:Detection}

\subsection{Parallel polarization}

Depending on the polarization of the subradiant Dicke states, 
the two ${}^{88}$Sr atoms will either repel (parallel polarization) 
or attract (perpendicular polarization) each other. To detect the 
atoms and infer the presence of polarization- and entanglement-dependent 
interactions, we propose below suitable detection protocols for 
both attractive and repulsive dynamics.
Both schemes rely on fluorescence detection methods, 
which turn out to be more advantageous as compared to 
schemes relying on ionization. 

\begin{figure*}[t!]
\centering
\includegraphics[width=0.9\linewidth]{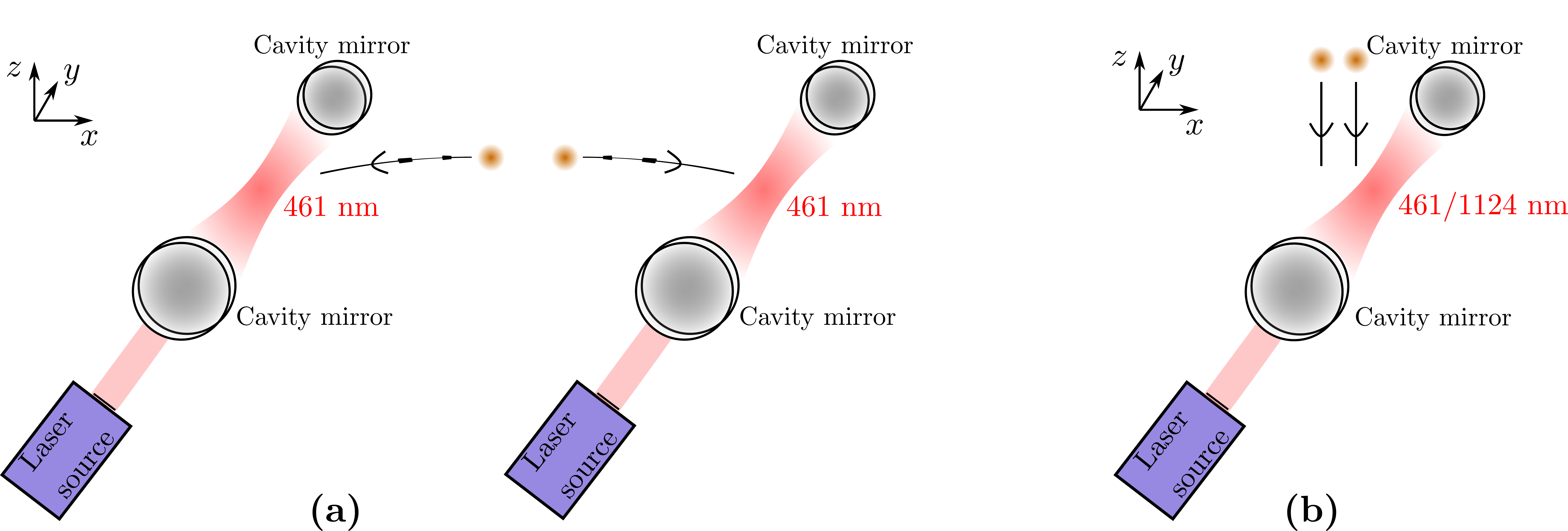}
\caption{\label{fig5} 
Sketch of the experimental apparatuses that can be employed to detect the
subradiant Dicke states with (a) parallel and (b) perpendicular polarization
with respect to the internuclear axis.  Both detection schemes rely on
fluorescence imaging after a certain time of flight. Fluorescence is induced by
cavities whose wavelength is resonant with either the blue MOT or the
$5{}^1P_1 \to 6{}^1S_0$ transitions of ${}^{88}$Sr.  }
\end{figure*}

In the case of repulsive interactions, the two atoms move in opposite 
directions along the $x$ axis, see also Fig.~\ref{Fig:Wave_prop}(a).
A suitable detection method would be to place two ``speed trap'' cavities 
in which laser-driven fluorescence will be induced
for both atoms [Fig.~\ref{fig5}(a)]. 
The location of the cavities and their wavelength can be 
determined by taking into account the following two considerations.

First, when the atoms fly away from each other, their interatomic 
distance becomes much larger than the wavelength of 
the ${}^{88}$Sr blue MOT transition ($\lambda_0 = 460.86\,\rm{nm}$).
This implies that the collective decay width is now 
comparable to the natural decay width~\cite{Jentschura_atoms_2023}, 
$\Gamma_0$, and therefore both atoms quickly decay to their ground electronic state.
To detect the atoms we can consider two laser cavities, whose 
wavelengths are in resonance with $\lambda_0$, therefore 
exciting the blue MOT line for each individual atom.
Such lasers are experimentally available~\cite{laser461}.
The laser intensity needs to be high in order to 
produce an appreciable Rabi frequency, and hence a strong fluorescence signal.
For the blue transition, we use the magnitude of the reduced electric-dipole
matrix element $\rdme{5\,{}^1S_0}{5\,{}^1P_1}=5.252(6)\,ea_0$
(see Table~6 of Ref.~\cite{PuKrKr88Sr}).
Relying on the relevant reduced matrix 
elements~\cite{WeGrThGa1992,PuKrKr88Sr,SaNa2010,NISTASD_2024,PuEtAl2025}, 
a laser of intensity, $I_L$, results in a Rabi frequency 
$\Omega(5{}^1S_0 \to 5{}^1P_1) = 2\pi \times 0.1065 \sqrt{I_L/(\rm{W}/\rm{cm}^2)}\,\rm{GHz}$.
With a beam waist of $w=10~\rm{\mu m}$, a cavity enhancement 
factor of $\mathcal{F}=100$, and a laser power of $P=40~\rm{mW}$, 
the intensity reads $I_L = 1.27 \times 10^6\,\rm{W}/\rm{cm}^2$, 
resulting in $\Omega(5{}^1S_0 \to 5{}^1P_1) = 2\pi \times 120~\rm{GHz}$.
This Rabi frequency is much larger than the natural decay width,
$\Gamma_0/\hbar = 2\pi \times 30.298(69)~\rm{MHz}$. While the detection scheme
assumes that atoms have already decayed to the ground state, it is reassuring
to observe that at least $600$ Rabi cycles can be induced before radiative
decay of the $5P$ level, so that spontaneous decay does not play a role in the
quantum dynamics governing the detection scheme. Hence, fluorescence detection
of the subradiant Dicke states with parallel polarization should be
experimentally feasible.

The second factor we need to consider for positioning the
``speed trap'' cavities is the gravitational sag that the
${}^{88}$Sr atoms experience during their time of flight.
Initially, the atoms are placed at $R= x_0 = 50\,\rm{nm}$.
The potential energy at the reference point is  
$E_{\rm pot}=V^{(\parallel)}_-(x_0) = h \times 173.0\,\rm{MHz}$.
This converts to a terminal velocity in the laboratory frame of 
approximately 
$v_{\infty} = \frac{1}{2} \sqrt{2E_{\rm pot}/\mu} = 0.8863 ~ \rm{m/s}$.
Under the given kinematic conditions, the gravitational sag is small.
Indeed, if we place the two ``speed trap'' cavities $5\,\rm{cm}$ 
away from the initial position of the atoms, the gravitational 
sag amounts to a small distance of $z_0 = 1.56\,\rm{cm}$ (using $g=9.81\,\mathrm{m/s^2}$) [see also Fig.~\ref{fig5}(a)].
Essentially, the two ${}^{88}$Sr atoms move in 
opposite directions in an almost horizontal trajectory.

\subsection{Perpendicular polarization}

For the subradiant Dicke states with transverse polarization,
one can devise two possible detection methods.
Both of these are based on irradiation in 
a resonant cavity centered at or lying below the 
spatial region in which the entangled state was
initially prepared [see also Fig.~\ref{fig5}(b)].

The first method is similar to the one employed 
for the states with parallel polarization.
Specifically, a cavity is placed far from the initial atomic configuration, 
along the $z$ axis, and the wavelength is resonant with the blue MOT 
transition of individual ${}^{88}$Sr atoms.
The two atoms will come close together owing to the attractive 
interactions [see also Fig.~\ref{Fig:Wave_prop}(b)], 
but eventually will fall down due to the gravitational pull.
Since the collective Dicke state is characterized by an 
incomplete suppression of the atomic natural decay width, 
the atoms are subject to decay toward  their electronic ground states.

For an initial interatomic distance $R=x_0 = 50\,\rm{nm}$, the 
lifetime of the subradiant Dicke state with perpendicular 
polarization is approximately~\cite{Jentschura_atoms_2023} 
$\tau^{\perp}_- = 57.9\,\rm{ns}$,
which corresponds to a tenfold enhancement
as compared to the natural lifetime.
Therefore, if we place the cavity at any distance 
substantially larger 
than $16~\rm{fm}$ below the initial atomic configuration,
then both atoms will have already decayed to their ground states.
As argued in the preceding section, reasonable Rabi frequencies 
can be derived for the driving of the blue MOT transition, 
leading to enhanced fluorescence signals. This detection scheme works after
decay to the ground state.

For the attractive state, a second approach is available to induce resonance
fluorescence before decay to the ground state. The other scheme relies on the
immediate detection 
of atoms right after the release from the subwavelength optical lattice.
In this regime, the atoms are still prepared 
in the entangled subradiant Dicke state.
The cavity is now resonant with the $5{}^1P_1 \to 6{}^1S_0$ 
transition, characterized by the wavelength $\lambda_0 = 1124.4~\rm{nm}$. 
Such a cavity induces a laser-driven dipole transition from the
subradiant entangled $5S$--$5P$ state
(denoted as $| \epsilon_P \rangle$)
to an entangled $5S$---$6S$ state 
(dubbed $| \epsilon_S \rangle$), 
\begin{subequations}
\begin{align}
| \epsilon_P \rangle = & \; \tfrac{1}{\sqrt{2}} \left(
| (5{}^1S_0)_A \, (5{}^1P_1)_B \rangle - | (5{}^1P_1)_A \, (5{}^1S_0)_B \rangle \right) \,,
\\
| \epsilon_S \rangle = & \; \tfrac{1}{\sqrt{2}} \left(
| (5{}^1S_0)_A \, (6{}^1S_0)_B \rangle - | (6{}^1S_0)_A \, (5{}^1S_0)_B \rangle \right) \,.
\end{align}
\end{subequations}

The interaction Hamiltonian of the two atoms with 
the fluorescence-inducing laser  is
\begin{equation}
H_I = -e (z_A + z_B ) E_\LL(t),
\end{equation}
where $E_\LL(t) = E_\LL \cos(\omega_\LL t)$. 
Here, we assume that the laser field is polarized along $z$, while  
$z_A$ and $z_B$ are the $z$ coordinates of the active
electrons in the identical atoms $A$ and $B$, 
and $E_L(t)$ is taken in the dipole approximation.
The transition matrix element is
\begin{align}
\langle \epsilon_P | H_I | \epsilon_S \rangle &= 
-\frac{e}{2} \biggl\{ \langle (5{}^1P_1)_A | z_A | (6{}^1S_0)_A \rangle  \nonumber \\
&+ \langle (5{}^1P_1)_B | z_B | (6{}^1S_0)_B \rangle \biggr\}
\, E_L(t)  \nonumber \\
&= -e \langle 5{}^1P_1 | z | 6{}^1S_0 \rangle \, E_L(t) \,,
\end{align}
which is equal to the corresponding transition matrix 
element for the $5{}^1P_1 \to 6{}^1S_0$ transition in an
individual ${}^{88}$Sr atom.
For this numerical estimate, we use the magnitude of the reduced
electric-dipole matrix element
$\rdme{5\,{}^1P_1}{6\,{}^1S_0}\simeq3.61\,ea_0$,
obtained from the theoretical transition probability
$A(6\,{}^1S_0\to5\,{}^1P_1)=18.6\times10^6\,\mathrm{s^{-1}}$
given by Werij \textit{et al.} (see Table~III of Ref.~\cite{WeGrThGa1992}).
Expressing the electric field in terms of the laser intensity, 
$I_\LL$, namely  $I_\LL = \frac12 \, \epsilon_0 \, c \, E_\LL^2$,
one obtains the Rabi frequency, 
$\Omega(5{}^1P_1 \to 6{}^1S_0) = 2\pi \times 73.2 \sqrt{I_\LL/ \rm{W/cm^2}}~ \rm{MHz}$.
Employing the same laser intensity as in the previous section, 
one gets $\Omega(5{}^1P_1 \to 6{}^1S_0) = 2\pi \times 82~ \rm{GHz}$.
Still, many Rabi cycles can be induced before the spontaneous radiative decay, 
and thus the fluorescence signal is strong enough to allow for the atom detection.
Moreover, we briefly note that lasers addressing the $5{}^1P_1 \to 6{}^1S_0$ 
transition are readily available~\cite{AuEtAl2002}.

\section{Conclusion}
\label{Sec:Conclusions}

We have calculated analytically the resonant dipole-dipole interatomic
potential inducing strong coupling between two atoms prepared in electronic
(super- or subradiant) Dicke states. The sign of these interactions is
controllably tuned by the polarization of the Dicke states relative to the
internuclear axis. We argue that selective preparation of long-lived
subradiant Dicke states, featuring a strong suppression of their collective
decay width, facilitates the experimental detection of these parametrically
enhanced, resonant interactions.

To demonstrate the polarization dependence of these interactions in
${}^{88}$Sr atoms, a two-step protocol was proposed here. One needs to prepare
the alkaline-earth-metal atom pair in a state where the nuclei are about
50~nm apart (as outlined in Appendix~\ref{appb}). Here a promising pathway
involves subwavelength optical lattices. However, for a first measurement, one
might also consider elongated optical dipole traps. The preparation of such
pairs is challenging, but within the reach of ultracold-atom experiments.
Releasing the atoms from the trap triggers the nontrivial relative interatomic
wave-packet propagation, which captures the impact of the resonant dipole-dipole
interactions.

We showed that Dicke states with a polarization parallel to the internuclear
axis experience a drastic increase of the mean position of the wave packet in
the course of the evolution, in view of a strong repulsive interaction
potential. In contrast, transverse polarization results in the progressive
fragmentation and dynamical trapping of the wave packet within short
interatomic distances. The oscillatory beats of the atoms' mean position are
associated with the population of quasimolecular bound states and can be tuned
as a function of the wave packet's mean initial position. Finally, we aim to
enrich our knowledge of bonding and antibonding in long-range entangled
alkaline-earth-metal dimers, at the single-molecule level. The interatomic
forces for the long-lived ``dark'' states are studied directly, and these
forces act for a sufficiently long time to make detection in technically
feasible ``speed trap'' cavities.

Based on our results, it would be intriguing to explore the effect of resonant
dipole-dipole interactions between identical atoms, for arbitrarily polarized
Dicke states, granting access to anisotropic, polarization, and
entanglement-dependent dipole-dipole interactions. These would generate a
complex landscape of unconventional molecular states. The entanglement-
dependent nature of the interaction cannot be understood based on classical
physics and represents a pure quantum phenomenon. Our investigations have
implications in new phases of matter that are currently explored with ultracold
polar molecules~\cite{bigagli_observation_2024,yuan_extreme_2025}.

\section*{Acknowledgments}

The authors acknowledge insightful conversations with Professor William D. Phillips. This work was supported by the National Science Foundation through
Grants No. PHY-2110294 and No. PHY-2513220 (U.D.J.). S.I.M. acknowledges
support from the Army Research Office under Award No. W911NF-26-1-A043.

\section*{Data Availability}

The data that support the findings of this article are not publicly available
upon publication because it is not technically feasible and/or the cost of
preparing, depositing, and hosting the data would be prohibitive within the
terms of this research project. The data are available from the authors upon
reasonable request.

\appendix

\section{Molecular Quantum Numbers}
\label{appa}

The preparation scheme for the long-range alkaline-earth-metal dimers
outlined in the present article is 
very much inspired by the notion of entangled Dicke states.
In the past, long-range alkali-metal dimers and alkaline-earth-metal
dimers have been studied 
in the context of photoassociation in traps (see Refs.~\cite{Jones2006,McAlexander_radiative_1996,Jones_measurement_1996,Nagel_photoassociative_2005,Escobar_two_2008}).
In the latter context, a different language has been used
for their description, namely, the quantum numbers
assigned to molecular states of dimers
coming from chemistry.
The purpose of this appendix is to find the translation between the entanglement-driven ``physics''
language and the ``chemical'' quantum numbers of the 
entangled states. It turns out that the 
long-range $\Sr(5s^2\,{}^1S_0)+\Sr(5s5p\,{}^1P_1)$
dimer provides an unusually clean
so-called Rosetta Stone for translating between three descriptions of the same 
one-excitation Hilbert
space, namely, {\em (i)} entangled Dicke states, {\em (ii)} Hund's case (a) and 
{\em (iii)} Hund's case (c).

In the language of quantum optics, one forms symmetric and antisymmetric Dicke
states from superpositions originating from the two possible 
lattice sites, or, in a more general sense, locations of the excitation. In Hund's case~(a),
the projection $\Lambda$ of the electronic orbital angular momentum [see also Eq.~\eqref{eq:casea}] and the
electronic-spin projection $\Sigmas$ (onto the same axis) label the molecular state. 
In Hund's case~(c), the corresponding label is the projection $\Omega$ of the total electronic
angular momentum $\bm J$. 
For the pure singlet manifold composed of the singlet $S$ and $P$ 
states of  $^{88}$Sr, $S=0$ implies $\Sigmas=0$, $J=L$, and $\Omega=\Lambda$. 
The nuclear spin $I=0$ independently removes hyperfine splitting. 

First, let us recall that the entangled Dicke states studied here are not globally the lowest molecular states of every symmetry.
Namely, the absolute electronic ground state is $X\,{}^1\Sigma_g^+$,
where, according to common notation in chemistry, 
$X$ denotes the molecular ground state. In global
short-range spectroscopic ordering, the attractive ungerade
variant of the ground state, $^1\Sigma_u^+$,
still associated with the $X\,{}^1\Sigma_g^+$, is commonly 
denoted by $2\,{}^1\Sigma_u^+$.
Stein, Kn\"ockel, and Tiemann studied its minimum region by Fourier-transform
spectroscopy \cite{Stein2008}. In the long-range photoassociation
of alkaline-earth-metal dimers, however, the
asymptotic symmetry label $^1\Sigma_u^+$ is usually offered as a 
sufficiently precise notation~\cite{Nagel_photoassociative_2005,Jones2006}.
Here, $g$ and $u$ refer to the {\em gerade} and {\rm ungerade} 
orbitals, corresponding to orbitals which are 
symmetric and antisymmetric (change sign) under an
inversion through the molecular center.

For the $SP$ entangled states, the total molecular electron angular momentum is $J=1$. 
This can be seen as follows. 
First, one observes that, for the 
$^1S_0$ state, the total electronic orbital angular momentum is $L = 0$, 
the total electronic spin-angular momentum is $S = 0$, 
and, for an $S$-state atom involved in the long-range binding, 
one has the total atomic electronic angular momentum $J_a = 0$.
Second, for the 
$^1P_1$ state, the total electronic orbital angular momentum is $L = 1$, 
the total electronic spin-angular momentum is $S = 0$, 
and, for a $P$-state atom involved in the long-range binding, 
one has the total atomic electronic angular momentum $J_a = 1$.
Hence, the total molecular electron angular momentum is $J=1$. 

Here, 
we assume that the internuclear axis $\bm R=\bm R_A-\bm R_B$
defines the quantization axis, which coincides with the $x$ axis.
For each spherical component $q=0,\pm1$, one defines
\begin{equation}
\ket{q}_B \equiv \ket{g}_A \, \ket{e,q}_B,
\qquad
\ket{q}_A \equiv \ket{e,q}_A \ket{g}_B,
\label{eq:localized}
\end{equation}
where the ground state is the singlet--$S$ state, and the 
excited state is the singlet--$P$ state.
The case $q=0$ corresponds to our longitudinally polarized
state, while the two perpendicularly polarized states
are linear combinations of the states with $q = \pm 1$.

Accordingly, the ``bright'' (superradiant) and ``dark'' 
(subradiant) Dicke states are the linear combinations,
\begin{equation}
\ket{q,\pm}=\frac{\ket{q}_B\pm\ket{q}_A}{\sqrt2}.
\label{eq:DickePM}
\end{equation}
They are the two-atom, one-excitation Dicke/Bell states in the $q$ polarization
sector \cite{Dicke_coherence_1954,Lehmberg1970II,Jentschura_atoms_2023}.
For an $S+P$ asymptote, inversion through the molecular midpoint, modeled by the operator $\Iop$, interchanges the atomic centers
$A \leftrightarrow B$ and contributes an odd intrinsic parity of the $P$ orbital. 
In a consistent conventional phase choice,
\begin{equation}
\Iop\ket{q}_A = -\ket{q}_B ,
\qquad
\Iop\ket{q}_B = -\ket{q}_A.
\label{eq:inversionlocalized}
\end{equation}
It follows that the superradiant Dicke state
\begin{equation}
\Iop \ket{q,+} = -\ket{q,+} \,,
\qquad
\ket{q,+} = \ket{q;u} \,,
\end{equation}
is an {\em ungerade} orbital, while 
the subradiant Dicke state,
\begin{equation}
\Iop \ket{q,-} = \ket{q,-} \,,
\qquad
\ket{q,-} = \ket{q;g} \,,
\end{equation}
is a {\em gerade} orbital.
The exchange-symmetric Dicke combination is therefore \emph{ungerade}, not
gerade, in this conventional $S+P$ phase convention.

In the language of Hund's case (a),
one organizes the electronic state by projections on the 
internuclear axis. One defines
\begin{equation}
\Lambda=\frac{\bm L\cdot\Rhat}{\hbar},
\qquad
\Sigmas=\frac{\bm S\cdot\Rhat}{\hbar},
\qquad
\Omega=\Lambda+\Sigmas.
\label{eq:casea}
\end{equation}
The familiar spectroscopic notation is
${}^{2S+1}\Lambda_{g/u}^{\pm}$.
Here, the Greek letter records $|\Lambda|$,
with an atomic-physics inspired notation where
$|\Lambda|=0,1,2,3,\ldots$ corresponds to 
$\Sigma,\Pi,\Delta,\Phi,\ldots$ orbitals.
The superscript $2S+1$ is the spin multiplicity. 
For the present manifold of singlet $S$ and singlet $P$ states,
one has, universally $S=0$, $\Sigmas=0$,
and $\Omega=\Lambda$.
For a $\Sigma$ state, the final superscript $+$ or $-$ is the 
eigenvalue under reflection through a plane containing 
the internuclear axis, which is a manifestly 
distinct operation compared to the inversion through the molecular center.
One observes that reflection through any plane 
containing the molecular axis $\bm R$ leaves 
a $P$ orbital with zero angular-momentum projection 
onto the molecular axis unchanged. Hence, 
both ${}^1\Sigma_g$ and ${}^1\Sigma_u$ states 
have positive reflection parity and qualify themselves
as ${}^1\Sigma_g^+$ and 
${}^1\Sigma_u^+$ states.

For transverse polarization, 
the subradiant Dicke 
state qualifies itself ($S=0$) as the 
{\em gerade} ${}^1\Pi_g$ states,
while the superradiant Dicke 
state qualifies itself ($S=0$) as the 
{\em ungerade} ${}^1\Pi_u$ state.
It is easy to show 
that reflection through a plane containing 
the internuclear axis interchanges the 
$q=+1$ and $q=-1$ components. Hence, 
no independent $+$ or $-$ superscript is 
attached to the bare electronic $\Pi$ term.  
The two inversion partners are written
as ${}^1\Pi_g$ for the subradiant state,
and ${}^1\Pi_u$ for the superradiant one.

In the language of Hund's case (c), orbital and spin angular momenta 
are first combined into the total electronic angular momentum $\bm J$,
and the electronic state is organized by its body-fixed projection $\Omega$,
\begin{equation}
\bm J=\bm L+\bm S, \qquad
\Omega=\frac{\bm J\cdot\Rhat}{\hbar}.
\end{equation}
The standard homonuclear labels are $\Omega_{g/u}^{\pm}$,
where $\Omega$ is a nonnegative integer.
The reflection superscript is required as an 
independent label for $\Omega=0$.
Case~(c) is normally the more appropriate language when spin--orbit 
coupling is strong enough that $\Lambda$ and $\Sigmas$ are 
not individually good quantum numbers, while $\Omega$ remains meaningful.  

For our spinless alkaline-earth-metal dimer,
the case (a) and case (c) labels contain the same 
information in the ideal $^1S_0+{}^1P_1$ manifold.
We can finally give a {\em Rosetta--Stone} dictionary to 
translate the Dicke-state labels into Hund's case (a) 
and Hund's case (c) labels of the molecular bound states,
\begin{subequations}
\label{rosetta}
\begin{align}
\ket{\Psi^{(\parallel)}_-} & = \ket{{}^1\Sigma_g^+} = \ket{0_g^+}, 
& & \mbox{(subradiant, repulsive)}  \nonumber \\ \\
\ket{\Psi^{(\parallel)}_+} & = \ket{{}^1\Sigma_u^+ } = \ket{0_u^+},
& & \mbox{(superradiant, attractive)} \nonumber \\ \\
\ket{\Psi^{(\perp)}_-}     & =  \ket{{}^1\Pi_g} = \ket{1_g},
& & \mbox{(subradiant, attractive)} \nonumber \\ \\
\ket{\Psi^{(\perp)}_+}     & = \ket{{}^1\Pi_u}  =\ket{1_u}.
& & \mbox{(superradiant, repulsive)} \nonumber \\
\end{align}
\end{subequations}
Here, in the chemical language, a {\em repulsive} state corresponds to an {\em antibonding} orbital,
while an {\em attractive} state is a {\em bonding} orbital.
The short-range asymptotics for $\Sigma$ states are, in absolute
magnitude, twice as large ($\pm 2 C / R^3$) compared to 
the short-range asymptotics for $\Pi$ states ($\pm C / R^3$).

\section{Trap Architectures}
\label{appb}

The preparation of a single $^{88}\mathrm{Sr}$ atom pair at the short length
scales considered here is experimentally challenging. As already highlighted in
the main text, one possibility is to prepare two ground-state ${}^{88}$Sr atoms
in subwavelength optical lattices.  Here, we further explore alternative
trapping schemes for bringing two atoms close together.

A first tempting idea might be to consider two tight optical
tweezers~\cite{Kaufman_quantum_2021,Norcia2018,Urech2022} placed side by side
at a distance of $50~ \rm{nm}$.  However, a closer inspection reveals that the
two tweezers would have to be very tightly focused, otherwise the two traps
would merge into a single one, obstructing the controllability of the initial
wave packet. In fact, the waist of a tweezer trap is limited by the
employed laser wavelength and the numerical aperture. Therefore, the waist cannot be
smaller than a few hundred nanometers, which is far larger than the
distances employed in our investigations.

Another possibility is to consider an elongated optical dipole
trap~\cite{Grimm2000}.  The atoms there reside in a Gaussian wave packet with
maximum density at zero interatomic distance.  Such a configuration can be
useful in the case of attractive subradiant Dicke states (perpendicular
polarization), where the resonant dipole-dipole interatomic potential can
sustain a multitude of bound quasimolecular states.  In particular, a
transition can be induced by means of photoassociation~\cite{Jones2006} to
specific bound states possessing a favorable Franck-Condon overlap with the
initial wave packet.  The small average interatomic distance will thus be
determined by the density profile of the final quasimolecular state.  A
problem associated with this approach is that it is also probable to excite
simultaneously quasimolecular levels pertaining to the bright counterpart,
{\em i.e.}, the superradiant, attractive Dicke states. Employing the notation
from Appendix~\ref{appa}, the relevant state is the ${}^1\Sigma_u^+$ attractive
interatomic potential.  Nevertheless, at $50\,\rm{nm}$, the subradiant state is
well resolved since the transverse subradiant-superradiant splitting is roughly
$122\,\mathrm{MHz}$, compared with a subradiant radiative width pertaining to
the perpendicular polarization of approximately $2.8\,\mathrm{MHz}$.  Moreover, this method does not apply to
repulsive Dicke states, since it relies on photoassociation to bound levels.

Finally, it is also possible to initialize the atoms in a five-state Raman
lattice, with the hope of granting a subwavelength spacing of $50~\rm{nm}$.  In
particular, employing the metastable $^3P_2$ state of ${}^{88}$Sr atoms, with
five Zeeman components, one can create a cyclic Raman lattice with interleaved
minima at a distance of $\lambda/(2N)$, where $\lambda$ is the laser wavelength
and $N$ is the number of Zeeman levels.  Utilizing the ${}^3P_2 \to {}^3D_2$
transition~\cite{Stellmer2014}, with $\lambda=496.93 ~\rm{nm}$, a subwavelength
lattice can be constructed with an effective spacing of $d=49.69 ~ \rm{nm}$.
This scheme relies on a recent experimental proposal for engineering qudits
with ${}^{88}$Sr atoms~\cite{Heizenreder_engineering_2026}.  Moreover, a
similar scheme has been successfully implemented in ${}^{87}$Rb
atoms~\cite{Anderson2020}, resulting in a three-level Raman subwavelength
lattice.  Despite its appeal, this method prepares atoms in a different
electronic manifold (${}^3P_2$) compared to the target electronic state
(${}^1P_1$).  A subsequent coherent transfer must therefore preserve the
prepared external wave functions, after which a separate $461\,\mathrm{nm}$
excitation is required to reach the ${}^{1}S_{0}+{}^{1}P_{1}$ manifold.
Overall, the elongated dipole trap provides a simpler setup, whereas the
subwavelength Raman lattice offers more direct control of the initial
interatomic separation.

\end{document}